\documentclass[10pt,journal,compsoc]{IEEEtran}
\usepackage{soul,color,xcolor}
\usepackage{graphicx}
\usepackage{subfigure}
\usepackage{amsmath,amssymb,amsfonts,amsthm}
\usepackage[ruled,linesnumbered]{algorithm2e}
\usepackage{cite}
\SetCommentSty{itshape} 
\usepackage{multirow}
\usepackage{tabularx}
\usepackage{booktabs}
\begin{document}

\title{ALANINE: A Novel Decentralized Personalized Federated Learning For Heterogeneous LEO Satellite Constellation}
\author{Liang Zhao,~\IEEEmembership{Member,~IEEE,}
        Shenglin Geng,
        Xiongyan Tang,~\IEEEmembership{Senior Member, IEEE,}
        Ammar Hawbani,
        Yunhe Sun,
        Lexi Xu, ~\IEEEmembership{Senior Member, IEEE,}
        Daniele Tarchi,~\IEEEmembership{Senior Member, IEEE,}
\IEEEcompsocitemizethanks{\IEEEcompsocthanksitem Liang Zhao, Shenglin Geng, Ammar Hawbani and Yunhe Sun are with the School of Computer Science, Shenyang Aerospace University, Shenyang 110136, China (e-mail: lzhao@sau.edu.cn; shenglin\_geng@yeah.net; anmande@ustc.edu.cn; sunyunhe@email.sau.edu.cn)}
\IEEEcompsocitemizethanks{\IEEEcompsocthanksitem Xiongyan Tang and Lexi Xu are with Research Institute, China United Network Communications Corporation, Beijing 100048, China (e-mail: tangxy@chinaunicom.cn; davidlexi@hotmail.com)}
\IEEEcompsocitemizethanks{\IEEEcompsocthanksitem Daniele Tarchi is with the School of Electronic and Information Engineering, University of Bologna, 40126 Bologna, Italy (e-mail: daniele.tarchi@unifi.it)}
\IEEEcompsocitemizethanks{\IEEEcompsocthanksitem Ammar Hawbani and Yunhe Sun are the corresponding authors}}

\IEEEtitleabstractindextext{%

\begin{abstract}
Low Earth Orbit (LEO) satellite constellations have seen significant growth and functional enhancement in recent years, which integrates various capabilities like communication, navigation, and remote sensing. However, the heterogeneity of data collected by different satellites and the problems of efficient inter-satellite collaborative computation pose significant obstacles to realizing the potential of these constellations. Existing approaches struggle with data heterogeneity, varing image resolutions, and the need for efficient on-orbit model training. To address these challenges, we propose a novel decentralized PFL framework, namely, \underline{A} Novel Decentra\underline{L}ized Person\underline{A}lized Federated Learning for Heteroge\underline{N}eous LEO Satell\underline{I}te Co\underline{N}st\underline{E}llation (ALANINE). ALANINE incorporates decentralized FL (DFL) for satellite image Super Resolution (SR), which enhances input data quality. Then it utilizes PFL to implement a personalized approach that accounts for unique characteristics of satellite data. In addition, the framework employs advanced model pruning to optimize model complexity and transmission efficiency. The framework enables efficient data acquisition and processing while improving the accuracy of PFL image processing models. Simulation results demonstrate that ALANINE exhibits superior performance in on-orbit training of SR and PFL image processing models compared to traditional centralized approaches. This novel method shows significant improvements in data acquisition efficiency, process accuracy, and model adaptability to local satellite conditions.

\end{abstract}

}

\maketitle

\IEEEdisplaynontitleabstractindextext

\IEEEpeerreviewmaketitle

\section{Introduction}
\IEEEPARstart Over the past few decades, there has been a significant increase in the number of satellites launched into Low Earth Orbit (LEO), along with a considerable improvement in the complexity of their functions. Initially dedicated to specific tasks such as communication, navigation, and remote sensing, modern LEO satellites now integrate these functions to provide comprehensive services \cite{1}. Additionally, the expansion in satellite numbers within constellations and improvements in computational capabilities have substantially boosted their on-orbit computational power, which enables more complex data processing tasks. Despite these advancements, managing the variability in data types and volumes collected by different satellites poses a considerable obstacle. The development of efficient inter-satellite collaborative computation methods is essential to fully utilize on-orbit computational resources for real-time data processing, which remains a critical bottleneck in realizing the potential of LEO satellite constellations \cite{3}. This integration of diverse functionalities necessitates innovative solutions to synchronize and harmonize the operations among satellites, which ensures that data relay and processing can be achieved seamlessly across the constellation.

In the context of heterogeneous LEO satellite constellations, characterized by differences in LEO satellite types and orbits, the obstacles previously discussed become even more pronounced \cite{4}. The diversity in satellite specifications and orbital paths introduces new complexities in the coordination and scheduling of task transmissions. Typically, satellites engage in data exchange among themselves for the purpose of data processing. However, the increase in data volumes combined with the high velocity of satellite orbits results in compromised data transmission quality. Frequent data transfers exacerbate the situation, which leads to intense competition for limited satellite communication bandwidth \cite{6}. This competition not only degrades the quality of data-processing results but also increases the latency in communication, further complicating real-time data analysis tasks. 

Efficient data processing emerges as a critical concern within the satellite constellation, and Federated Learning (FL) offers a promising solution as an innovative form of distributed processing technology \cite{7}. By processing data locally on each satellite, FL necessitates only the transmission of model parameters, rather than raw data. This approach not only significantly reduces the volume of data transferred, but also enhances process efficiency and protects against potential data theft and tampering during transmission. Nevertheless, challenges arise within the framework of heterogeneous LEO satellite constellations. Traditional FL can lead to severe communication congestion and intense bandwidth competition between satellites and servers, which may adversely affect the overall performance of Machine Learning (ML) models. To address these issues, decentralized FL (DFL) supports dynamic and direct interactions between satellites, thus alleviating communication bottlenecks \cite{10}.

Furthermore, the variability in data quality collected by different satellites complicates the application of these data in local model training and global model aggregation, often leading to suboptimal model accuracy \cite{11}. The deployment of a globally trained model to individual satellites for further local training may not always yield optimal results. This suboptimal performance can occur when the model fails to align adequately with the specific characteristics of local datasets, potentially leading to ineffective localized training effects \cite{12}. To mitigate the issue of uneven data distribution, personalized FL (PFL) allows each satellite to employ a personalized model tailored to its local data distribution, rather than strictly adhering to a global model \cite{13}. This adaptation enables more effective and efficient learning results tailored to specific local conditions.

To address the challenges and build upon the motivations discussed in the previous discussions, we propose ALANINE, a novel decentralized PFL framework specifically designed for heterogeneous LEO satellite constellations. ALANINE employs image Super Resolution (SR) to enhance the quality of data acquired by satellites and utilizes DFL to facilitate on-orbit training of the SR model. Furthermore, it leverages inter-satellite links to transmit the trained SR model. Each satellite applies the trained SR model to improve the quality of its local data before engaging in PFL. Subsequently, decentralized PFL methods are used for training FL image processing models. Model pruning is employed to tailor the global model to the specific conditions of each satellite local dataset, thus enhancing the adaptability and usability of the global model in local training. ALANINE enables training of on-orbit DFL satellite image SR and PFL image processing models, which achieves efficient data acquisition and processing while improving the accuracy of PFL image processing models. The main contributions of this paper are as follows.

\begin{itemize}
\item {Considering the data heterogeneity of LEO satellites, we propose a novel decentralized PFL framework for heterogeneous LEO satellite constellation, specifically designed to manage the unique data heterogeneity of each satellite. This approach allows for the efficient aggregation and personalized training of local models, significantly improving the integration and effectiveness of global models.}

\item To address the varing image resolutions that affect the training accuracy of satellite-based models, we implement a DFL method for satellite image SR. This innovative method significantly enhances the resolution of input data, thus directly improving the training accuracy, reliability, and robustness of models under varing image quality conditions, crucial for accurate remote sensing analysis.

\item To enhance the training efficiency of PFL image processing models in satellite constellations, our approach employs advanced model pruning during both the local training and global aggregation phases. This strategy effectively reduces model complexity, increases transmission efficiency, and allows quicker and more adaptive responses to the unique challenges of local satellite environments. These optimizations are crucial for maintaining high-performance PFL models in the resource-constrained and demanding contexts of satellite constellations.

\end{itemize}

{The rest of this paper is organized as follows:} In Section \ref{sec_rel_work}, we introduce the related work. Section \ref{sec_sys_model} presents the system model. Section \ref{sec_pro_sol} describes a DFL approach for satellite image SR, utilizes model pruning to enhance model efficiency, and develops a decentralized PFL architecture. The performance of our framework is evaluated in Section \ref{pre_eva}. Section \ref{conclusion} concludes the work.

\begin{figure}[t]
\centering
\includegraphics[width=0.5\textwidth]{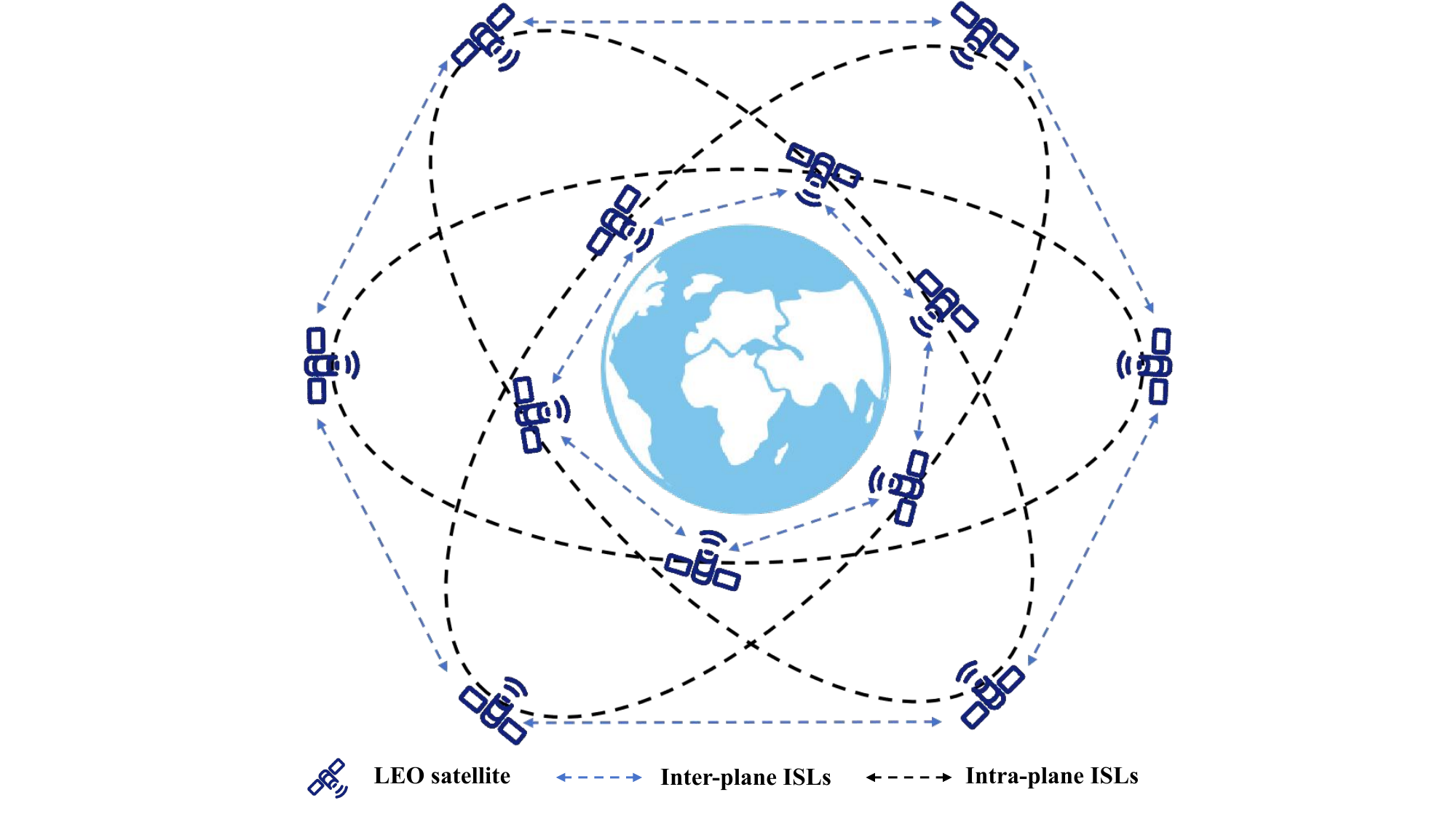}
\caption{Illustration of the Walker Star satellite constellation.}
\label{fig:1}
\end{figure}

\section{{Related Work}} {\label{sec_rel_work}}

This section begins with an overview of recent advances in satellite-based FL. Subsequently, it reviews various studies that explore the application of SR techniques in satellite image processing. An in-depth analysis of model pruning within the framework of FL is then discussed. Finally, the section highlights the significance and potential impacts of the proposed ALANINE framework, which emphasizes its noteworthy contributions.

\subsection{Satellite Federated Learning}

FL in satellite networks, which transmits only updated model parameters to a parameter server, significantly reduces communication costs and improves data utilization. Recent studies have investigated the feasibility and efficiency of this implementation in satellite networks. Razmi $\mathit{et ~ al.}$ \cite{FL1} propose a method that uses Inter-Satellite Links (ISLs) within LEO constellations to enhance learning efficiency and reduce communication loads through asynchronous FL. Complementing this approach, Han $\mathit{et ~ al.}$ \cite{FL2} propose a ground-to-satellite cooperative FL using LEO satellites as Edge Computing (EC) units and model aggregators to enhance ML services in remote areas. However, the process of achieving synchronous and effective model updates presents challenges due to infrequent satellite access to the parameter server.

To mitigate the challenges associated with frequent communication with the parameter server, the DFL method is introduced. Yan $\mathit{et ~ al.}$ \cite{FL3} optimize the convergence time in DFL by designing the number of LEO satellites and the number of orbits. Zhai $\mathit{et ~ al.}$ \cite{FL4} propose an offloading-assisted DFL framework, for LEO satellite networks. This framework enhances model training by leveraging inter-satellite collaboration to optimize both latency and accuracy. Furthermore, PFL has demonstrated effectiveness in addressing the discrepancies between local and global models, as well as handling the issue of data heterogeneity. In \cite{FL5}, a PFL framework integrated with Model-Contrastive Learning (PFL-MCL) is proposed, effectively addressing the challenges of large-scale heterogeneous multi-modal data in a human-centric Metaverse. By employing multi-center aggregation and a multi-modal fusion network, the framework significantly enhances the efficiency of personalized training and communication in model deployment.

\subsection{Satellite Image Super Resolution}

SR plays a crucial role in enhancing the resolution and quality of images and videos by reconstructing high-resolution outputs from their low-resolution counterparts \cite{SR1}. Recent developments have leveraged Convolutional Neural Networks (CNNs), and Generative Adversarial Networks (GANs) to address challenges associated with upscaling artifacts and to achieve more realistic SR images. In \cite{SR2}, an Edge-Guided Video SR framework is proposed by integrating the Multi-Frame SR model with the Edge Single-Frame SR model within a unified network, which aims to enhance the SR performance of video satellite imagery. Furthering these advancements, Ni $\mathit{et ~ al.}$ \cite{SR3} introduce a novel continuous-scale satellite video SR network. This network incorporates a residual-guided, time-aware dynamic routing deformable alignment module along with scale-aware feature extraction, effectively enhancing feature alignment accuracy and scale adaptability for improved resolution results.

In addition to leveraging DL for SR, the introduction of GANs into satellite image SR marks a significant advancement, which showcases the maturation and enhancement of this approach. For instance, Karwowska $\mathit{et ~ al.}$ \cite{SR4} innovatively enhances the Enhanced SRGAN (ESRGAN) by integrating a multi-column discriminator and employing Wasserstein loss, significantly improving the speed and quality of SR generation for satellite images. Zhao $\mathit{et ~ al.}$ \cite{SR5} present a spatial fusion method named SRSF-GAN, which utilizes GAN and SR module. This method only requires low-resolution image at the time of prediction and high-resolution reference image selected based on spatial-spectral similarity criteria, effectively achieving efficient spatial fusion of satellite images. 

\begin{table}[t]
	\caption{ Notations}
	\begin{center}
		\resizebox{8.5cm}{6cm}{
		\begin{tabular}{l|l}
			\hline
			\textbf{Notation} & \textbf{Description}     \\   \hline                                     
            $N$ & The total number of satellites \\ 
            $G$ & The total number of orbits \\
            $S$ & The number of satellites per orbit \\    
            $s_i$ & Satellite ID \\   
            $s_g^s$ & Satellite ID within each orbit $g$\\  
            $d \theta (s_i,s_j)$ & The maximum inclination range between $s_i$ and $s_j$ \\
            $h_{s_i}$, $h_{s_j}$ & The orbital heights of $s_i$ and $s_j$ respectively \\
            $r_d$ & The Earth radius \\
            $r_T$ & The effective communication range threshold \\
            $d(s_i,s_j,t)$ & The distance between $s_i$ and $s_j$ at time $t$ \\
            $R_i$ & The transmission rate of $s_i$ \\
            $B$ & The bandwidth of the channel \\
            $p_{i}^{C}$ & The transmission power of $s_i$ \\
            $g_{i}$ & The channel gain of $s_i$ \\
            $N_{p}$ & The noise power spectral density \\
            $\mathcal{D}$ & The data set \\
            $\omega$ & The model parameter vector \\
            $R$ & The number of FL rounds\\
            $\kappa$ & The number of local training epochs\\
            $F(\omega)$ & The global loss function \\
            $F_i(\omega)$ & The local loss function for $s_i$\\
            $f(X,\omega)$ & The loss function for each individual sample $X$ \\
            $\mathbf{L}$ & The low-resolution image \\
            $\mathbf{H}$ & The high-resolution image \\
            $F_{map}(\cdot)$ & The mapping function \\
            $K_1, K_2, K_3$ & The filters\\
            $k_1, k_2, k_3$ & The spatial dimensions of the filters \\
            $b_1, b_2, b_3$ & The biases \\
            $C$ & The number of input image channels \\
            $M$ & The number of satellites participating in the SR model training \\
            $\phi$ & The SR model parameters \\
            $\eta$ & The learning rate \\
            $\mathbf{J}$ & The adjacency matrix \\
            $\mathcal{S}_n^t$ & The neighborhood set of $s_n$ at time $t$ \\
            $\mathbf{h}$ & The mask \\
            $\odot$ & The Hadamard product \\
            $\varepsilon_c, \varepsilon_v$ & The threshold \\
            $c_t(\cdot)$ & The voting function \\
            $p_f$ & The pruning frequency \\
            $\alpha$ & The exponential parameter \\
            $\beta$ & Parameters for the delayed optimal initial pruning time \\
            $\gamma$ & The scaling factor \\
            $\hat{\mu}$ & The sparsity \\
            \hline   
		\end{tabular}}
        \label{table_sys_model}
	\end{center}
\end{table}

\subsection{Model Pruning for Federated Learning}

In FL, model pruning optimizes network efficiency by removing redundant parameters, significantly reducing computational demands without major performance loss. Recent studies show that pruning not only lessens computational load but also combats overfitting, thus improving model generalization. Liu $\mathit{et ~ al.}$ \cite{MP1} propose a joint model pruning and device selection method for wireless FL, which optimizes the pruning ratio, device selection, and wireless resource allocation to enhance learning performance and reduce communication overhead. Furthermore, it introduces a threshold-based device selection strategy. Du $\mathit{et ~ al.}$ \cite{MP2} introduce a novel FL framework that significantly reduces computational and communication cost in mechanical fault diagnosis. This framework achieves efficiency gains through streamlined training processes, non-structural model pruning, and fine-tuning, complemented by an optimal model selection strategy.

Moreover, Jiang $\mathit{et ~ al.}$ \cite{MP3} propose a pioneering FL approach named PruneFL, which utilizes adaptive and distributed parameter pruning to notably decrease communication and computation overhead on edge devices while preserving accuracy levels comparable to the original model. Following a similar enhancement strategy, Jiang $\mathit{et ~ al.}$ \cite{MP4} unveil an efficient FL framework that improves computational and communication efficiency across heterogeneous nodes via adaptive model pruning. Additionally, Zhou $\mathit{et ~ al.}$ \cite{MP5} develop the novel FL framework FedPAGE, designed specifically to tackle global efficiency challenges in heterogeneous environments with the use of dynamic adaptive pruning.

\subsection{Summary}

As demonstrated in the aforementioned literature, the heterogeneity of data from LEO satellites poses significant challenges to the efficient training and aggregation of FL models in satellite constellations. These challenges often result in poor performance of localized training when applying the global model to individual satellites. Furthermore, current research has not addressed the quality of data collected by satellites, nor the potential for optimizing training efficiency through model pruning during the FL training process. To address these issues, we introduce ALANINE, a decentralized PFL framework designed for heterogeneous LEO satellite constellation. ALANINE aims to resolve the data heterogeneity issues in FL training by adopting PFL strategies and model pruning, thus enhancing the efficiency of model training. Furthermore, ALANINE employs DFL satellite image SR to improve the quality of data collected by satellites, and the trained SR model are transmitted to each satellite to be applied in the preprocessing stage of PFL image processing models. This system not only ensures the privacy of data quality and model parameters, but also enhances the accuracy and practicality of the FL models.

\section{SYSTEM MODEL}{\label{sec_sys_model}}
In this section, the network model integrated with the DFL framework is investigated, and a detailed analysis of the communication and computation models related to ALANINE is conducted. The primary symbols employed throughout this paper are summarized in Table \ref{table_sys_model}.
\subsection{Network Model}

The ALANINE framework is introduced to enhance the data quality collected by satellites and to improve the accuracy of PFL image processing models. As illustrated in Fig. \ref{fig:1}, our study is conducted within a Walker Star satellite constellation \cite{SM1}, which consists of $N$ satellites, arranged in $G$ orbits with $S$ satellites per orbit, leading to the relationship $N = G \times S$. Within this constellation, satellites are organized into sets denoted as $\mathcal{N} = \{1, 2, \ldots, N\}$. Each orbital plane $g \in \mathcal{G}$, part of $\mathcal{G} = \{1, 2, \ldots, G\}$, maintains a specific altitude above the Earth. $\mathcal{G}$ includes all orbital planes within the heterogeneous LEO satellite constellation. Each satellite is uniquely identified by an index $i$, represented as $s_i \in \{s_1, s_2, \ldots, s_N\}$. In every orbital plane $g$, satellites follow identical trajectories and are uniquely identified within their respective planes as $\mathcal{S}_g^{'} = \{s_g^1, s_g^2, \ldots, s_g^S\}$. This notation ensures that $s_g^{s+1}$ is positioned directly behind $s_g^s$ in the same orbital trajectory. Satellites on different orbits can communicate via ISLs to conduct training tasks for various FL models, thus enhancing the efficiency of data collection and processing across the satellite constellation.

\subsection{Communication Model}

We consider a scenario where each satellite is equipped with four communication devices. Two of these devices are designated for intra-orbital communication, which facilitates the aggregation of models within the same orbital plane. The remaining two devices are used for inter-orbital communication, which are essential for aggregating multiple intra-orbital models into a global model. For the ISLs, communication is feasible if the line of sight is not obstructed by the Earth. Current research indicates that at altitudes higher than 80 km above sea level, the quality of ISLs is not significantly degraded by atmospheric conditions \cite{SM2}. For satellites $s_i$ and $s_j$, the maximum inclination range $d \theta (s_i,s_j)$, which can be represented as:

\begin{equation}
d\Theta(s_i,s_j)=\sqrt{(h_{s_i}+r_d)^2-r_T^2}+\sqrt{(h_{s_j}+r_d)^2-r_T^2}
\end{equation}

\noindent where $h_{s_i}$ and $h_{s_j}$ represent the orbital heights of $s_i$ and $s_j$ respectively, $r_d$ is the Earth radius, and $r_T$ is the effective communication range threshold. This equation calculates the maximum permissible inclination for maintaining a clear line of sight between the two satellites, which ensures effective communication without Earth interference. We assume that communication between $s_i$ and $s_j$ becomes feasible when their distance at time $t$ is less than $d\Theta(s_i,s_j)$, which can be expressed as follows:

\begin{equation}
d(s_i,s_j,t) < d\Theta(s_i,s_j)
\end{equation}

\noindent where $d(s_i,s_j,t)$ denotes the distance between $s_i$ and $s_j$ at time $t$. This ensures that the satellites are within the effective communication range, which is critical for maintaining the integrity and reliability of the data exchange processes necessary for the operation of satellite constellations.

The data transmission rate $R_i$ for $s_i$, is calculated using Shannon formula, which is formulated as follows:

\begin{equation}
R_{i} =B\log_2(1+\frac{P_{i}^{C}g_{i}}{N_{p}B})
\end{equation}

\noindent where $B$ represents the bandwidth of the channel, $p_{i}^{C}$ denotes the transmission power of $s_{i}$, and $N_{p}$ is the noise power spectral density. The channel gain associated with $s_i$ is designated as $g_{i}$. This approach, which considers critical parameters significantly impacting the efficiency and reliability of communications, characterizes the dynamics of data transmission between satellites.

\subsection{Computation Model}
In a heterogeneous LEO satellite constellation, each satellite employs its inherent onboard computational capabilities to engage collaboratively in the training of a ML model. This collaborative endeavor utilizes datasets $\mathcal{D}$ that are uniquely available to each satellite. The initial ML model is pre-defined and uniformly distributed across all satellites. The primary goal is to solve an optimization problem:

\begin{equation}
\mathop{\arg\min}\limits_{\omega\in\mathbb{R}^d} F(\omega)= \frac{1}{|\mathcal{D}|} \sum\limits_{X\in \mathcal{D}}f(X,\omega)
\label{F(1)}
\end{equation}

\noindent where $\omega$ is defined as the model parameter vector. The global loss function is represented by $F(\omega)$, is critical to assessing the model effectiveness across the full satellite network. Here, $f(X,\omega)$ denotes the loss function for each individual sample $X$ in the dataset.

Due to the limited computational resources on each satellite and the challenges associated with data transmission between satellites, FL is employed for training. The optimization problem described by \eqref{F(1)} is solved iteratively using the Stochastic Gradient Descent (SGD) approach, wherein the optical objective function can be separable as:

\begin{equation}
\mathop{\arg\min}\limits_{\omega\in\mathbb{R}^d} F(\omega)= \sum\limits_{i\in \mathcal{N}}\frac{|\mathcal{D}_{i}|}{|\mathcal{D}|}F_{i}(\omega)
\label{f(1)}
\end{equation}

In FL, $s_i$ possesses a local dataset $\mathcal{D}_i$, such that $|\mathcal{D}|$ can be expressed as $|\mathcal{D}|=\sum\limits_{i=1}^N |\mathcal{D}_{i}|$, $F_{i}(\omega)$ is defined as the local loss function for $s_i$,  which can be denoted as:

\begin{equation}
F_{i}(\omega) = \frac{1}{|\mathcal{D}_{i}|}\sum\limits_{X \in \mathcal{D}_{i}} f_{i}(X, \omega)
\label{14}
\end{equation}

The optimization process differs from conventional FL approaches by adopting a decentralized FL method, where each satellite shares parameters with others to iteratively update and further optimize the global model parameters, ultimately achieving optimal results. The specific process is detailed in Section \ref{sec_pro_sol}.


\section{APPROACH DESIGN}{\label{sec_pro_sol}}

In this section, considering the varying image resolution captured by satellites, we propose a DFL for satellite image SR algorithm to enhance image resolution in \ref{DFL SR}. To improve the efficiency of local model training in PFL, we introduce a dynamic model pruning strategy for PFL in \ref{MP}. In response to the unique data distribution characteristics of satellites, we develop a decentralized PFL for heterogeneous LEO satellite constellation in \ref{DPFL}.

\begin{figure*}[t]
\centering
\includegraphics[width=0.9\linewidth]{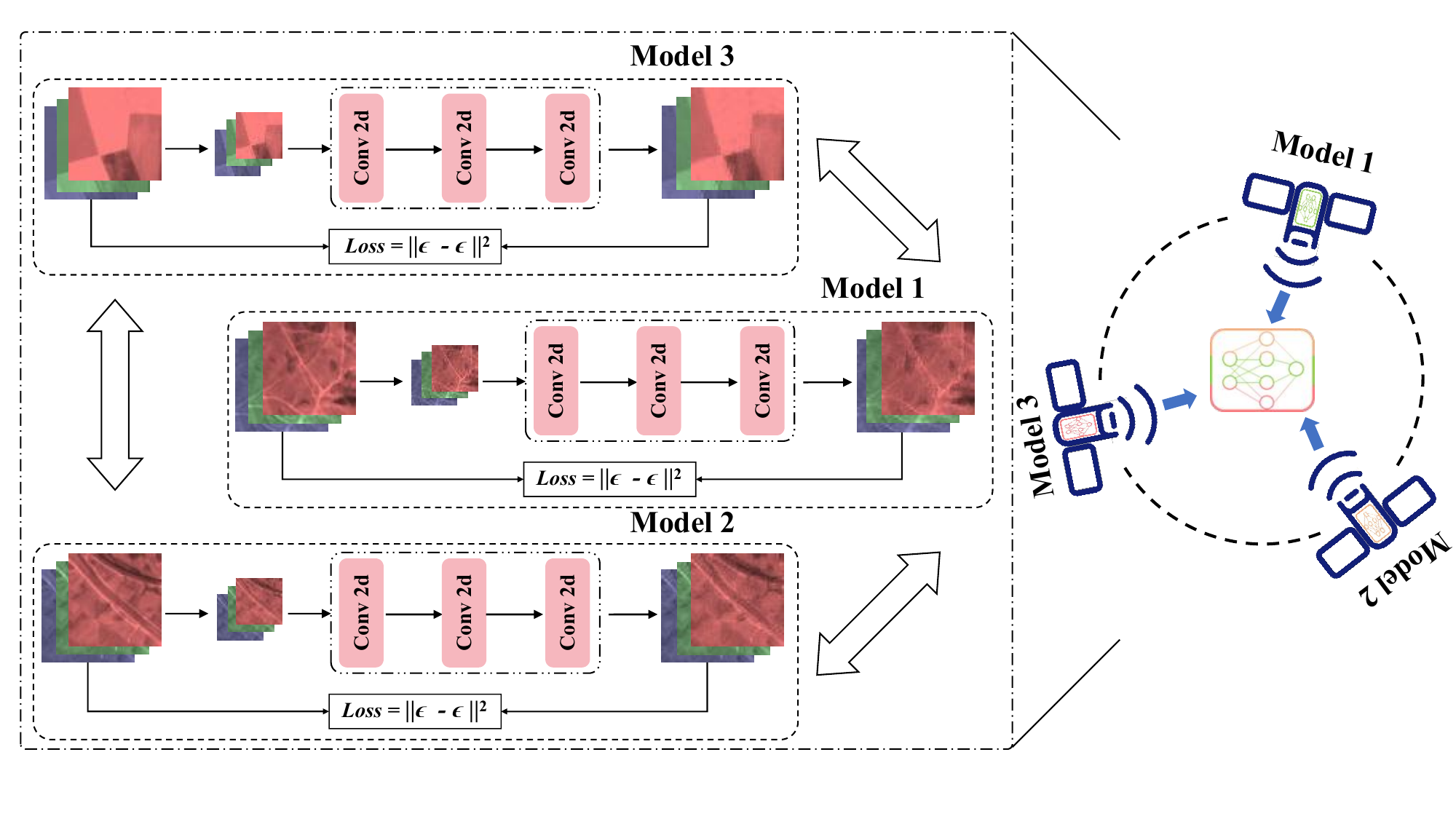}
\caption{Representation of decentralized federated learning super resolution in Walker Star satellite constellation.}
\label{fig:2}
\end{figure*}

\subsection{Decentralized Federated Learning for Satellite Image Super Resolution}\label{DFL SR}

The process of DFL for satellite image SR primarily utilizes CNN for model training. This process can be divided into three main components, which involves patch extraction and representation, non-linear mapping, and reconstruction. The overall architecture of DFL for satellite image SR is illustrated in Fig. \ref{fig:2}. The specific process of DFL for satellite image SR is outlined in Algorithm \ref{alg1}, where satellites are selected based on computational capacity. These satellites then perform local training and model updates using SGD, followed by global aggregation, which averages the weighted local models.

\subsubsection{Patch Extraction and Representation}

For low-resolution satellite images, we first perform a preprocessing operation on the low-resolution images using Bicubic interpolation. The interpolated image is represented as $\mathbf{L}$. Based on the low-resolution image $\mathbf{L}$ and the high-resolution image $\mathbf{H}$, we aim to obtain a mapping relationship $F_{map}$, which seek to generate an image $F_{map}(\mathbf{L})$ that closely approximates the high-resolution image $\mathbf{H}$. Patches are extracted from the low-resolution image $\mathbf{L}$, and each obtained patch is represented as a high-dimensional vector. Each vector encompasses a set of feature maps with multiple dimensions. Subsequently, joint optimization of the filters involved in the image convolution process and the network is implemented. The formula for this process can be represented as follows:

\begin{equation}
F_{map}^1(\mathbf{L})=\max\left(0,K_1*\mathbf{L}+b_1\right)
\end{equation}

\noindent where $F_{map}^1(\cdot)$ denotes the operation function of the first layer, $K_1$ represents the filters, and $b_1$ is defined as the biases. The number of input image channels is denoted by $C$, typically corresponding to the three color channels of an RGB image. The spatial dimensions of the filters are specified by $k_1$. The weight matrix $K_1$ performs $d_1$ independent convolution operations, each utilizing a kernel of size $C \times k_1 \times k_1$, which allows the network to extract multi-scale features. Additionally, the bias adjustments for the convolution outputs are managed by the vector $b_1$, which comprises $d_1$ elements, each aligned with a corresponding filter. Following the convolution operation, we employ the Rectified Linear Unit (ReLU) as the activation function. This function not only mitigates the vanishing gradient problem in CNN but also promotes sparse activation, thus enhancing the capacity of the model to learn more effective feature representations.

\begin{algorithm}[t]
    \caption{DFL for Satellite Image SR} 
    \label{alg1}
    \KwIn{initialize global model parameter $\phi_0$, maximum number of participating satellites $M$, set of all orbits $\mathcal{G}$}
    \KwOut{Updated global model parameter $\phi$}

    $\mathcal{M} \gets \{\}$ // Set of selected satellites participating in the training of the SR model\\

    \While{$|\mathcal{M}| < M$}{
    \For{each orbit $g \in \mathcal{G}$}{
        $s_g^i \gets \text{ComputationalCapacity}(s_g^j)$ // Select satellite with highest computational capacity in orbit $g$\\
        $\mathcal{M} \gets \mathcal{M} \cup \{s_g^i\}$\\
        \If{$|\mathcal{M}| \geq M$}{
            break;
            }
        }
    }
    \For{each FL round $ r \in \mathcal{R} $}{
        \For{each selected satellite $s_i \in \mathcal{M}$}{
            Compute local loss: $F_i(\phi) \gets \frac{1}{|\mathcal{D}_i|} \sum\limits_{j\in \mathcal{D}_i}||F_{map}(\mathbf{L}_{ij};\phi)-\mathbf{H}_{ij}||^2$ \\
            Update local model by \eqref{10} \\
        }
        
        Update global model by \eqref{f(1)} and \eqref{SGD} 
    }
    \Return{Final updated global model parameter $\phi$}
\end{algorithm}

\subsubsection{Non-linear Mapping}

Initially, $d_1$ dimensional features are extracted from each patch by the preceding layer. Subsequently, these $d_1$ dimensional vectors are mapped to $d_2$ dimensional vectors, and $d_2$ filters with dimensions of 1$\times$1 are applied. Through the non-linear mapping of each high-dimensional vector to another high-dimensional vector, a series of high-resolution patches is generated, which collectively form a new set of feature maps. The structure of the second layer resembles that of the previous layer, which aims to enhance the non-linear characteristics of the network model. The function of the second layer can be expressed as:

\begin{equation}
F_{map}^2(\mathbf{L})=\max\left(0,K_2*F_{map}^1(\mathbf{L})+b_2\right)
\end{equation}

\noindent where $F_{map}^2(\cdot)$ represents the operation function of the second layer, $K_2$ and $b_2$ correspond to the weights and biases of the filters, respectively. The weight matrix $K_2$ consists of $d_2$ filters, each with dimensions of $d_1 \times k_2 \times k_2$. The associated bias vector $b_2$ comprises $d_2$ elements. Each $d_2$ dimensional output vector represents a high-resolution patch, which can be employed in the image reconstruction process described in the \ref{Reconstruction}.

\subsubsection{Reconstruction}{\label{Reconstruction}}

In the reconstruction phase, the network leverages features extracted and mapped by preceding layers to reconstruct high-resolution output from the low-resolution input image. Specifically, the reconstruction layer employs a series of convolutional operations to transform high-dimensional feature maps into the final high-resolution image. Thus, the function of the convolutional layer can be denoted as:

\begin{equation}
F_{map}(\mathbf{L})=K_3*F_{map}^2(\mathbf{L})+b_3
\end{equation}

\noindent where $K_3$ denotes a set of $C$ filters, each with dimensions $d_2 \times k_3 \times k_3$, and $b_3$ represents the corresponding bias vector containing $C$ elements. Through the application of convolutional kernels and non-linear activation functions, the network generates high-resolution images with enhanced visual clarity and enriched details, effectively improving both image quality and resolution.

In our proposed DFL for satellite image SR, we aim to learn the mapping function $F_{map}$ by minimizing the loss between the generated images and their high-resolution counterparts. The network model parameters $\mathbf{\phi} $ are defined as $\mathbf{\phi} = \{K_1, K_2, K_3, b_1, b_2, b_3\}$. The loss is quantified using the Mean Squared Error (MSE) method, which can be defined as follows:

\begin{equation}
F(\phi)= \frac{1}{|\mathcal{D}|} \sum\limits_{i\in \mathcal{M}}\sum\limits_{j\in \mathcal{D}_i}||F_{map}(\mathbf{L}_{ij};\phi)-\mathbf{H}_{ij}||^2
\label{10}
\end{equation}

\noindent where $\mathcal{M}$ denotes the set of satellites participating in the training of the SR model, $\mathbf{L}_{ij}$ and $\mathbf{H}_{ij}$ represent the input low-resolution image and its corresponding high-resolution image from $s_i$, respectively. This equation  captures the multi-source nature of the dataset, utilizing imagery from various satellites to enhance the robustness and generalizability of the SR model. To minimize the aforementioned loss function, we also employ the SGD optimization method. The corresponding optimization function can be presented as follows:

\begin{equation}
\phi_i \leftarrow \phi - \eta \nabla_{\phi}F_i(\phi)
\label{SGD}
\end{equation}

\noindent where $\phi_i$ denotes the SR model parameters for $s_i$, and $\phi$ represents the locally trained SR model parameters.  Additionally, $\eta$ signifies the learning rate. The gradient of the local loss function with respect to the parameters $\phi$ is denoted by $\nabla_{\phi}F_i(\phi)$.


\begin{algorithm}[t]
    \caption{Dynamic Model Pruning Policy for PFL}
    \label{DP}
    \KwIn{model parameters $\omega_{n,\kappa}^t$, momentum $\mathbf{M}$, mask $\mathbf{h}_n^t$; norms $0 < s \leq 1 < j$, compression $\eta_m$, scaling factor $\delta$, prune threshold $\zeta$, prune rate $p_m$, epochs $\kappa$, pruning iterations set $\mathcal{T}$, number of batches $B$}
    \KwOut{model parameters $\hat{\omega}_{n,\kappa}^{t}$ and mask $\mathbf{h}_n^{t}$}

    \For{epoch $i \in \{1, \ldots, \kappa\}$}{
        \For{batch $b \in \{1, \ldots, B\}$}{
            $G \gets \nabla_{\omega_{n,i}^t} F(\omega_{n,i}^t, b)$ \\
            $\mathbf{M}, \omega_{n,i}^t \gets \text{UpdateParameters}(\mathbf{M}, \omega_{n,i}^t, G)$ \\
            $\omega_{n,i}^t \gets \omega_{n,i}^t \odot \mathbf{h}_n^t$
        }
        \For{$t \in \mathcal{T}$}{
            $M_{\text{total}} \leftarrow \sum_{n=1}^N \|\mathbf{M}_{n,i}\|_1$ \\
            $\Omega_{\text{pruned}} \leftarrow \sum_{n=1}^N |\{w \in \omega_{n,i}^t : |w| < p_m\}|$ \\
            \For{layer $l \in K$}{
                Compute model parameters: $d_l^t \gets |\mathbf{h}_n^{l,t}|$ \\
                PQ Index: $I_l(\omega_{n,\kappa}^{l,t}) \gets 1 - (\frac{1}{d_l^t})^{\frac{1}{j}-\frac{1}{s}} \frac{\|\omega_{n,\kappa}^{l,t}\|_s}{\|\omega_{n,\kappa}^{l,t}\|_j}$ \\
                The lower bound of model parameters: $r_l^t \gets d_l^t(1+\eta_m)^{-\frac{j}{j-s}} (1 - I_l(\omega_{n,\kappa}^{l,t}))^{\frac{j}{j-s}}$ \\
                The number of model parameters after prune: $c_l^t \gets \lfloor d_l^t \cdot \min(\delta(1-\frac{r_l^t}{d_l^t}), \zeta) \rfloor$ \\
                $m_l \leftarrow \text{getMomentum}(\mathbf{M}_l, \mathbf{h}_n^{l,t}, M_{\text{total}})$ \\
                $\text{prune}(\omega_{n,i}^{l,t}, \mathbf{h}_n^{l,t}, p_m)$ \\
                $\text{regrow}(\omega_{n,i}^{l,t}, \mathbf{h}_n^{l,t}, m_l \cdot \Omega_{\text{pruned}})$ \\
                $p_m \leftarrow \text{decayRate}(p_m)$ \\
                $\omega_{n,i}^{l,t} \gets \omega_{n,i}^{l,t} \odot \mathbf{h}_n^{l,t}$
            }
        }
    }

    \Return{$\hat{\omega}_{n,\kappa}^t, \mathbf{h}_n^t$}
\end{algorithm}

\subsection{Dynamic Model Pruning Policy for Personalized Federated Learning}\label{MP}

In order to enhance the efficiency of PFL in resource-constrained satellite environments, dynamic model pruning techniques are employed. The detailed process of this technique is elaborately outlined in Algorithm \ref{DP}. This algorithm not only describes the integration of local model updates with dynamic pruning but also incorporates pruning driven by the PQ Index and sparse momentum techniques. Such advanced methods are critical in managing the computational resources effectively, especially in the complex network of a satellite constellation.

In this paper, we consider a heterogeneous LEO satellite constellation comprising $N$ satellites uniformly distributed across $G$ orbital planes, with $S$ satellites per plane. These satellites can communicate both intra-plane and inter-plane through ISLs. The interconnectivity within the satellite constellation can be formally represented as $\mathcal{S}(N,\mathbf{J})$, where $N$ denotes the total number of satellites and $\mathbf{J}$ is the adjacency matrix. Specifically, $\mathbf{J}=[u_{nm}]\in \mathbb{R}^{N \times N}$ encapsulates the communication links between satellites, with each element $u_{nm}$ indicating the connectivity status between $s_n$ and $s_m$. For $s_n$, its set of neighboring satellite is denoted as $\mathcal{S}_n = \{m \in \mathcal{N} \mid u_{nm} > 0\}$. 

To implement a robust DFL method, we employ a dynamic communication strategy based on time intervals, where $t \in \mathcal{T}$. A time-varying symmetric network topology is defined through this approach. This topology is represented by the time-dependent set of adjacent matrix $\mathbf{J}^t=[u_{nm}^t]\in \mathbb{R}^{N \times N}$. Within this dynamic framework, the neighborhood $\mathcal{S}_n^t$ for $s_n$ at time interval $t$ is specified as $\mathcal{S}_n^t = \{m \in \mathcal{N} \mid u_{nm}^t > 0\}$. For $s_n$, the learning process involves training on its local dataset $\mathcal{D}_n$, and exchanging models with satellites in its neighborhood $\mathcal{S}_n^t$. Building upon \eqref{f(1)}, we can reformulate the optimization problem for PFL as follows:

\begin{equation}
\min_{\omega_{n},n\in[1,N]} F(\omega_{n})=\frac{1}{N}\sum_{n=1}^{N}F_{n}(\omega_{n})
\label{mask}
\end{equation}

\noindent where $F_n(\omega_n)=\mathbb{E}[f(X; \omega_n) |X \in \mathcal{D}_n)]$ defines the average loss function of the model on the local dataset $\mathcal{D}_k$. According to \eqref{mask}, we introduce a masking approach to implement model pruning within the DFL. This method aims to eliminate unnecessary model weights and reduce redundancy. Consequently, the PFL problem incorporating model pruning can be formulated as follows:

\begin{equation}
\min_{\omega, \mathbf{h}_n,n\in[1,N]} F(\omega_{n})=\frac{1}{N}\sum_{n=1}^{N}F_{n}(\omega\odot\mathbf{h}_{n})
\label{masking}
\end{equation}

\noindent where $\omega$ represents the global model, $\mathbf{h}$ denotes the mask, and $\odot$ symbolizes the Hadamard product. The function $F_n( \omega \odot \mathbf{h}_{n} )=\mathbb{E}[f(X; \omega \odot \mathbf{h}_{n}) |X \in \mathcal{D}_n)]$ represents the average loss of the pruned model on the local dataset $\mathcal{D}_n$. In \eqref{masking}, the objective is to determine an optimal global model $\omega$ and a set of individual masks $\mathbf{h}_n$ for $s_n$. This optimization process ensures effective communication between $s_n$ and its neighboring satellites within $\mathcal{S}_n^t$, while adhering to the constraints imposed by the time-varying adjacency matrix $\mathbf{J}^t$. This optimization process aims to derive personalized models for $s_n$ through the relation $\omega_n=\omega \odot \mathbf{h}_n$.

To address the limitations identified in \eqref{masking}, we propose a method that employs a mask to train on heterogeneous satellite data, while simultaneously improving model convergence time through dynamic network communication. This approach introduces a pruning mechanism at specific intervals during the training process, selectively retaining or removing model parameters based on their importance to the objectives of the satellite network. Given the orbital characteristics of heterogeneous LEO satellite constellation, a ring-connected topological structure is adopted for model training and transmission. 

In each round $t$ of PFL, $s_n$ is required to collect the locally trained models from its neighboring satellites defined by $\mathcal{S}_n^t$ and subsequently performs model aggregation. To initialize the masks, we leverage the Scale-Free Networks (SFN) distribution \cite{Nature} for sparse construction, which is complemented by a distributed implementation of the Sparse Momentum Algorithm (SMA) \cite{Neurisp} for mask removal and regrowth. The compression efficacy of the CNN model is assessed using the PQ Index \cite{PQI}, which is adapted for our decentralized PFL framework. Subsequently, a tailored pruning strategy is implemented, which considers the unique data distribution characteristics of $s_n$, thus ensuring personalized model optimization. For determining the optimal pruning timing $t$ across the network of $N$ satellites, we establish a threshold $\varepsilon_c$ and a voting funciton $c_t(\cdot)$. This approach can be formalized as:

\begin{equation}
c_t(n)=\begin{cases}1&\text{if}|\frac{\|\omega^t-\omega^0\|^2-\|\omega^{t-1}-\omega^0\|^2}{\|\omega^1-\omega^0\|^2}|<\varepsilon_c,\\0&\text{otherwise.}\end{cases}
\label{ctn}
\end{equation}

Consequently, the initial pruning time can be expressed as: $\hat{t} = \min\{t : \frac{1}{N}\sum_{n=1}^{N}c_{t}(n) < \varepsilon_{v}\}$. The voting ratio threshold is denoted by $\varepsilon_v$. In addition to the initial pruning time, we define a pruning frequency $p_f$. This parameter determines the intervals between subsequent pruning operations following the initial pruning, thus facilitating a more dynamic and adaptive pruning process across the satellite network. The pruning frequency $p_f$ can be expressed as:

\begin{equation}
p_f = \left\lfloor\frac{(\hat{t}+\beta)^{\alpha}}{\gamma^{r-1}} + 0.5\right\rfloor, \quad r \in \{1,2,\ldots,R\}
\label{15}
\end{equation}

Here, $\alpha$ is an exponential parameter controlling the growth pattern of pruning intervals, $\beta$ is used to optimize the initial pruning time delay. The function of the parameter as a modulator for the pruning frequency is served by $\gamma$. Additionally, $r$ represents the FL round. Therefore, the set of pruning times $\mathcal{T}$ can be represented as:

\begin{equation}
\mathcal{T} = \{t_{\chi} = \sum_{r=1}^{\chi} p_f : \hat{t} < t_{\chi} < R\}, \quad \chi \in\{\mathbb{Z}_{\geq1}\}
\label{16}
\end{equation}

Following the aforementioned pruning operations, the aggregated model $\hat{\omega}_n^t$ of $s_n$ at time $t$, based on the generated mask $\mathbf{h}_n$ and the models from the neighbor set $S_k$, which can be expressed as:

\begin{equation}
\hat{\omega}_n^t = \frac{\sum_{j \in \mathcal{S}_{n+}^t}\omega_j^t}{\sum_{j \in \mathcal{S}_{n+}^t}\mathbf{h}_j^t} \odot \mathbf{h}_n^t
\end{equation}

\noindent where $\mathcal{S}_{n+}^t$ represents the set of neighboring satellites $ \mathcal{S}_{n}^t$ including $s_n$. Similar to the method described in (\ref{SGD}), which applies the SGD method over $\kappa$ local training epochs, the expression for $\hat{\omega}_n^t$ is derived as follows:

\begin{equation}
\hat{\omega}_{n,\kappa+1}^{t}=\hat{\omega}_{n,\kappa}^{t}-\eta(\nabla_{\hat{\omega}}F_n(\hat{\omega})\odot\mathbf{h}_{n}^{t})
\end{equation}


\begin{algorithm}[t]
    \caption{Decentralized PFL for Heterogeneous LEO Satellite
Constellation}
    \label{alg3}
    \SetKwData{and}{\textbf{and}}
    \KwIn{heterogeneous LEO satellite constellation, satellite set $\mathcal{N}$, number of rounds $R$, epochs $\kappa$, sparsity $\hat{\mu}$, PQI hyperparameters $s$, $j$, $\delta$, $\eta_m$ }
    \KwOut{personalized models $\{\hat{\omega}_{n,\kappa}^R\}_{n=1}^N$}
    Initialize $\{\omega_n^0\}_{n=1}^{N}, \mathcal{T} \gets \emptyset$\\
    $\phi_R \gets$ Execute Algorithm \ref{alg1} for $R$ rounds to obtain final SR model\\
    Distribute $\phi_R$ to each satellite $ s_n \in \mathcal{N}$\\
    
    \tcc{Image Initialization using derived SR model}
    \For{each satellite $s_n \in \mathcal{N}$}{
        $\mathcal{D}_n \gets$ Local dataset of $s_n$
        
        $\mathcal{D}_n^{SR} \gets \emptyset$
        
        \For{each low-resolution image $\mathbf{L}_i \in \mathcal{D}_n$}{
            $\hat{\mathbf{H}}_i \gets F_{map}(\mathbf{L}_i; \phi_R)$ 
            
            Apply SR model:
            $\mathcal{D}_n^{SR} \gets \mathcal{D}_{n,\kappa}^{SR} \cup \{\hat{\mathbf{H}}_i\}$
        }
    }
    
    \tcc{Dynamic Pruning and Model Personalization}
    \For{round r $\in$ $\mathcal{R}$}{
    \For{each satellite $s_n \in \mathcal{N}$}{
        Initialize personalized model: $\omega_n^r \gets \omega$ \\ 
        $\omega_n^r \gets$ AggregateModels($\{\omega_m^r | m \in \mathcal{S}_n^r\}$) \\
         $\hat{\omega}_n^r \gets$ MaskBasedAggregate($\omega_n^r$) \\
        $\hat{\omega}_{n,\kappa}^r \gets$ LocalTrain($\hat{\omega}_n^r$, $\kappa$ ) \\
        Compute $c_r(n)$ by \eqref{ctn} \\
        Broadcast($c_r(n)$)\\
        $\hat{t}  \gets \min\{t : \frac{1}{N}\sum_{n=1}^{N}c_{r}(m) < \varepsilon_{v}\}$\\
        \If{$r \in \mathcal{T}$ \and $\mu_k < \hat{\mu}$}{
        Execute Algorithm \ref{DP} to achieve personalized model $\hat{\omega}_{n,\kappa}^r$
        
        Update $\mu_k$
        }
        $\omega_n^{r+1} \gets \hat{\omega}_{n, \kappa}^{r}$
        }
        \If{$r = \hat{t}$}{
            Update $\mathcal{T}$ by \eqref{15} and \eqref{16}
        }
    }
    \Return{$\{\hat{\omega}_{n,\kappa}^R\}_{n=1}^N$}
\end{algorithm}

\subsection{Decentralized Personalized Federated Learning for Heterogeneous LEO Satellite Constellation}\label{DPFL}

Given the varying resolutions of remote sensing images acquired by satellites and the limited computational capabilities of satellites, we employ the method described in \ref{DFL SR} to train an image SR model. This model is then transmitted to all satellites. Preceding the classification model training process delineated in \ref{MP}, the SR model is applied to enhance the satellite imagery to a uniform, higher spatial resolution. Following this process, individual satellites execute personalized classification model training tailored to its specific data distribution. This approach aims to enhance model accuracy and improve the localization performance of the global model.

Algorithm \ref{alg3} delineates the comprehensive framework for decentralized PFL for heterogeneous LEO satellite constellation. The process commences with the initialization of personalized models $\{\omega^0_{n,\kappa}\}^N_{n=1}$ for each satellite in the constellation. Subsequently, it executes Algorithm \ref{alg1} to obtain an SR model $\phi_R$, which serves as the foundation for subsequent image processing operations across the satellite constellation. The core of Algorithm \ref{alg3} lies in its iterative execution of dynamic pruning and model personalization. This intricate process comprises multiple critical steps executed for each round $r$ and each satellite $s_n$. Initially, Algorithm \ref{alg3} aggregates models from adjacent satellites, which applies a mask-based aggregation method to refine the model. Local training is then conducted on each satellite, which utilizes its specific dataset to further personalize the model. 

Algorithm \ref{alg3} effectively incorporates Algorithm \ref{DP} to obtain personalized models, which ensures the adaptability and efficiency of the global model during local training. This iterative approach facilitates continuous optimization of models for individual satellites. These models adapt to the distinct operational parameters of each satellite while leveraging the dataset of the constellation. By integrating these components, the proposed approach adeptly manages the complex tasks of SR image processing and personalized model training across the diverse satellite network. The developed methodology not only enhances the individual performance of each satellite, but also elevates the collective capabilities of the heterogeneous LEO satellite constellation in handling intricated image processing and ML tasks. Through the implementation of the described framework, our system substantially enhances its capacity to address the distinct requirements of individual satellites within the heterogeneous satellite constellation. This enhanced capability results in a more robust and efficient distributed learning framework for satellite constellations.

\begin{table}[t]
 \renewcommand{\arraystretch}{1.0}
 \caption{\centering Simulation parameters}
 \label{tab2}
 \begin{center}{
  \begin{tabular}{|l|l|l|}
   \cline{1-3}
   \textbf{Parameter} & \textbf{Symbol} & \textbf{Value} \\ \cline{1-3}
   The orbital altitude & $ H $ & 330km \\
   The inclination of the orbital plane & $ \phi $ & $90^{\circ}$ \\
   The constellation size & $ N $ & 50 \\
   The quantity of orbital planes & $ G $ & 5 \\
   The number of satellites per orbital plane & $ S_g^{'} $ & 10 \\
   The number of participating satellites & $ M $ & 5 \\ 
   The FL round & $ R $ & 100 \\
   The local epoch count & $ \kappa $ & 1,5 \\
   The allocated bandwidth & $ B $ & 20MHz \\
   The transmission power & $ p^{trans} $ & 5W \\
   The computation power & $ p^{comp} $ & 5W \\ 
   The satellite channel gain & $ g_i $ & 1000 \\
   The learning rate & $ lr $ & 0.01 \\
   The batch size & $ bs $ & 64 \\
   The voting ratio threshold & $ \varepsilon_v $ & 0.5 \\
   The norm index & $p$, $q$ & 0.5, 1 \\
   The scaling factor & $\delta$ & 0.9 \\
   The compression & $ \eta_m $ & 1 \\
   The Dirichlet parameter & $ \alpha $ & 0.3, 3 \\
 \cline{1-3}
  \end{tabular}
  }
\end{center}
\end{table}

\begin{table*}
\centering
\caption{Comparative Analysis of PSNR and SSIM Metrics for EuroSAT dataset}
\label{tab3}
\setlength{\tabcolsep}{4pt}
\begin{tabularx}{\textwidth}{@{}lc*{5}{>{\centering\arraybackslash}X}@{}}
\toprule
Metric & Scale & Bicubic & KK \cite{KK} & SRGAN \cite{SRGAN} & RDN \cite{RDN} & ALANINE \\
\midrule
\multirow{2}{*}{PSNR} & $\times$ 2 & 32.97 & 36.73 & 37.35 & 37.89 & \textbf{38.01} \\
 & $\times$ 4 & 29.93 & 31.62 & 31.79 & 31.99 & \textbf{32.08} \\
\midrule
\multirow{2}{*}{SSIM} & $\times$ 2 & 0.8754 & 0.9128 & 0.9390 & 0.9426 & \textbf{0.9491} \\
 & $\times$ 4 & 0.7683 & 0.8097 & 0.8184 & 0.8203 & \textbf{0.8215} \\
\bottomrule
\end{tabularx}
\end{table*}

\section{PERFORMANCE EVALUATION}{\label{pre_eva}}

This section commences with a delineation of the simulation scenario and specification of relevant parameter settings. Subsequently, the efficacy of the SR model is evaluated using multiple performance metrics. The classification accuracy of models trained on satellite imagery enhanced through the proposed DFL SR method is then assessed. Furthermore, we validate that ALANINE optimizes delay compared to different DFL methods. Finally, we summarize the results of the simulation and further highlight the advantages of our proposed solution.

\subsection{Evaluation Parameters and Simulation Scenarios}

The simulation of the LEO satellite constellation is implemented using the Walker Star constellation model in the Systems Tool Kit (STK)\footnote{STK is a product of AGI and serves as an advanced computational tool for the aerospace sector, facilitating detailed assessments of satellite trajectories and real-time position tracking}. The simulation platform utilized Python $3.9$ on a Windows $11$ $64$-bit operating system, which primarily employs the torch and matplotlib libraries for computational and visualization purposes. The source code has been made publicly available on Github\footnote{https://github.com/NetworkCommunication/ALANINE}. To facilitate the training of SR and satellite PFL models across the satellite network, we design a constellation comprising $50$ LEO satellites. These satellites are distributed across $5$ orbital planes, with $10$ satellites uniformly positioned in each orbit. The orbital configuration is characterized by an altitude of $330$ km above the surface of the Earth and an inclination angle of $90$ degrees. Table \ref{tab2} presents a detailed overview of the main simulation parameters.

For our simulation evaluation, we utilize the EuroSAT dataset\footnote{The EuroSAT dataset utilizes imagery obtained from the ESA Sentinel-2 mission, designed to systematically acquire optical imagery at high spatial resolution}, a comprehensive collection of satellite imagery specifically designed for land use and land cover classification tasks \cite{eurosat}. This dataset encompasses $10$ distinct classes, with a total of $27,000$ RGB images, each with spatial dimensions of $64\times64$ pixels. The dataset is split into training and test sets with a ratio of $8:2$. Our SR model comprises three convolutional layers, which are optimized for enhancing spatial resolution while preserving spectral information. Concurrently, we construct a PFL model consisting of four convolutional layers designed to effectively capture both global and client-specific features. To meet the requirements for training SR datasets, we construct low-resolution and high-resolution data forms by downsampling the original images. In order to rigorously evaluate the robustness of our models under realistic distributed learning scenarios, data partitioning strategies are implemented using the Dirichlet distribution method with two distinct parameters $(\alpha)$. This approach allows for the construction of different levels of data heterogeneity, which simulates varing real-world distributions.

\subsection{Evaluation of ALANINE Super Resolution Scheme}
To evaluate the performance of our proposed ALANINE SR scheme, extensive simulations are conducted using the EuroSAT dataset. The method is compared against several SR techniques including Bicubic interpolation, KK \cite{KK}, SRGAN \cite{SRGAN}, and RDN \cite{RDN}. We use Peak Signal-to-Noise Ratio (PSNR) and Structural Similarity Index (SSIM) as evaluation metrics. Table \ref{tab3} presents a comparative analysis for different upscaling factors ($\times 2$ and $\times 4$) across various methods. ALANINE consistently outperforms existing techniques in both PSNR and SSIM metrics across all scaling factors. For $\times 2$ upscaling, our method achieves a PSNR of $38.01$ dB and an SSIM of $0.9491$, surpassing RDN by $0.12$ dB and $0.0065$. These performance gains are consistent for $\times 4$ upscaling, where our method attains a PSNR of $32.08$ dB and an SSIM of $0.8215$, outperforming RDN by $0.09$ dB and $0.0012$, respectively. These results underscore the robustness of our ALANINE SR scheme across different scaling factors. The consistent superior performance in both PSNR and SSIM metrics suggests that our method not only achieves higher fidelity in pixel-wise reconstruction, but also preserves structural information more effectively than existing approaches. While the improvements may seem incremental in some cases, they represent significant advancements in the field of SR. These advancements potentially lead to noticeable improvements in visual quality for remote sensing application.

\begin{figure}[htbh]
    \centering
    \subfigure[]{
        \includegraphics[width=0.12\textwidth]{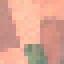}
    }
    \subfigure[]{
        \includegraphics[width=0.12\textwidth]{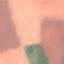}
    }
    \subfigure[]{
        \includegraphics[width=0.12\textwidth]{pic/KK.jpg}
    }
    \subfigure[]{
        \includegraphics[width=0.12\textwidth]{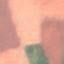}
    }
    \subfigure[]{
        \includegraphics[width=0.12\textwidth]{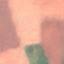}
    }
    \subfigure[]{
        \includegraphics[width=0.12\textwidth]{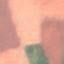}
    }
    \caption{Comparison of SR results across different algorithms. (a) Low-resolution input image; (b) Bicubic; (c) KK; (d) SRGAN; (e) RDN; (f) ALANINE.}
    \label{fig:A}
\end{figure}

Fig. \ref{fig:A} presents a qualitative comparison of $\times 4$ upscaling SR results obtained from various algorithms, which complements the quantitative analysis discussed previously. This comparison illustrates the performance of various methods on the dataset for $\times 4$ upscaling. The results demonstrate that our proposed method achieves improved reconstruction of fine textures, smoother color transitions, and better preservation of structural details, particularly in areas with subtle color variations. These visual enhancements corroborate the higher PSNR and SSIM values achieved by our method in SR.

\subsection{Evaluation of ALANINE Personalized Federated Learning Scheme}
To evaluate the performance of our proposed PFL method within the ALANINE framework, we conduct a series of detailed simulations, which compares it against several baseline algorithms. To ensure fairness in comparison, all methods are tested under identical simulation environments and parameters.

\begin{figure*} [htb]
 \begin{center}
 \subfigure[]{
  \label{fig1a} 
  \includegraphics[width=4.2cm]{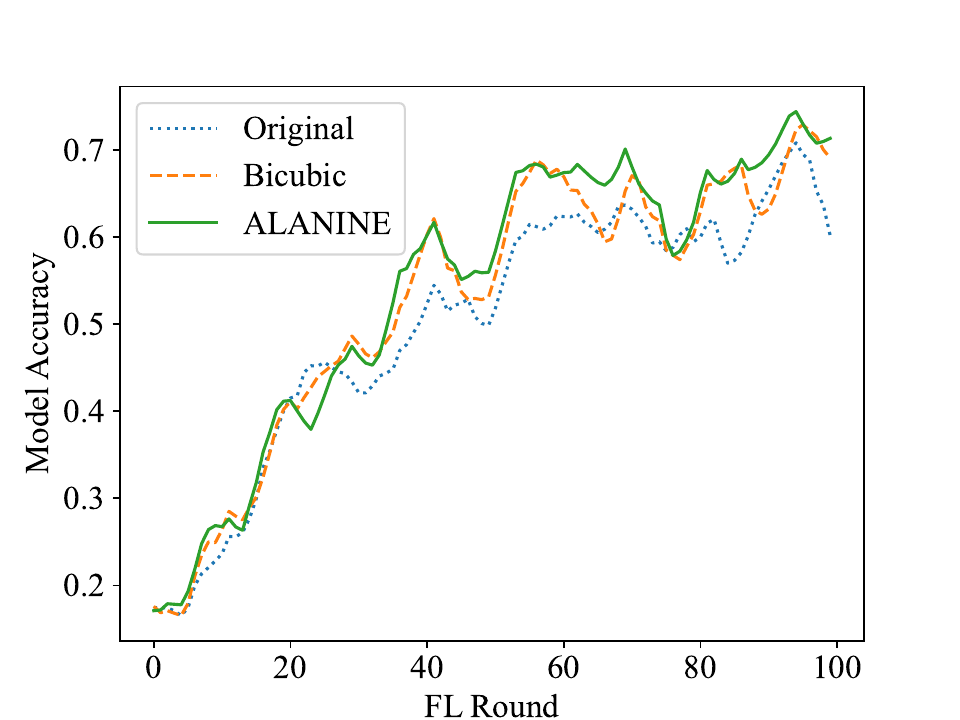}}
 \hspace{0in} 
 \subfigure[]{
  \label{fig1b} 
  \includegraphics[width=4.2cm]{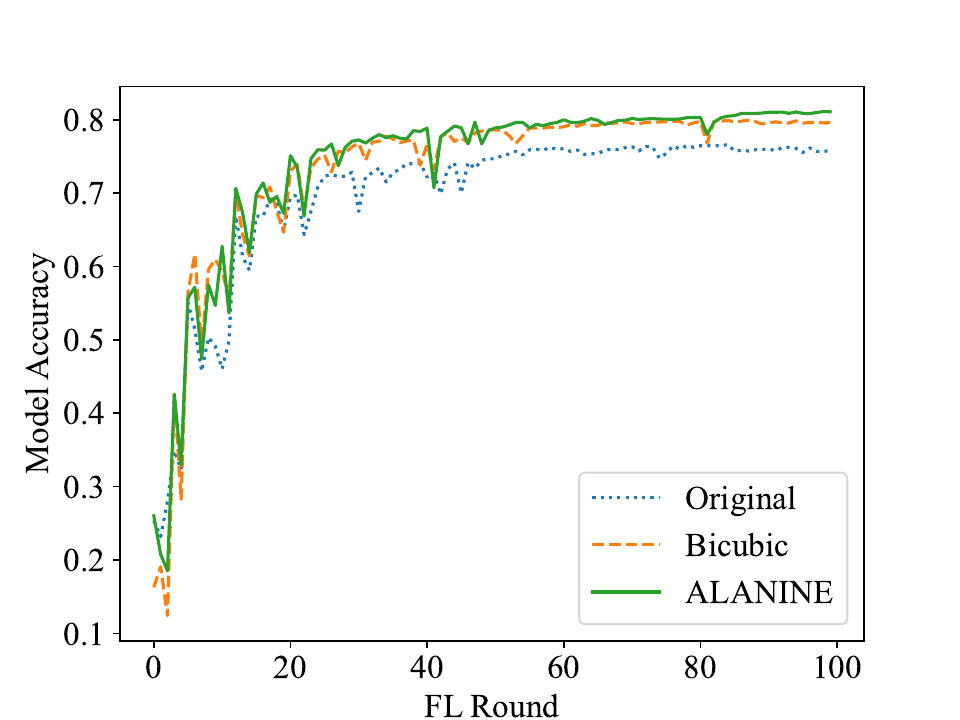}}
 \hspace{0in} 
 \subfigure[]{
  \label{fig1c} 
  \includegraphics[width=4.2cm]{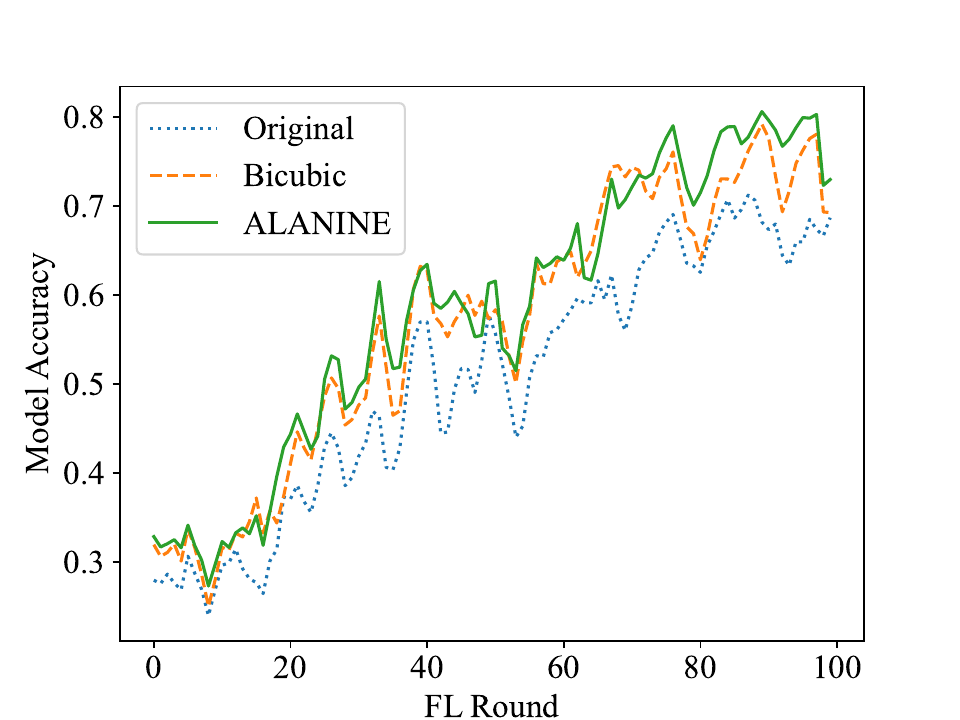}}
 \hspace{0in} 
  \subfigure[]{
  \label{fig1d} 
  \includegraphics[width=4.2cm]{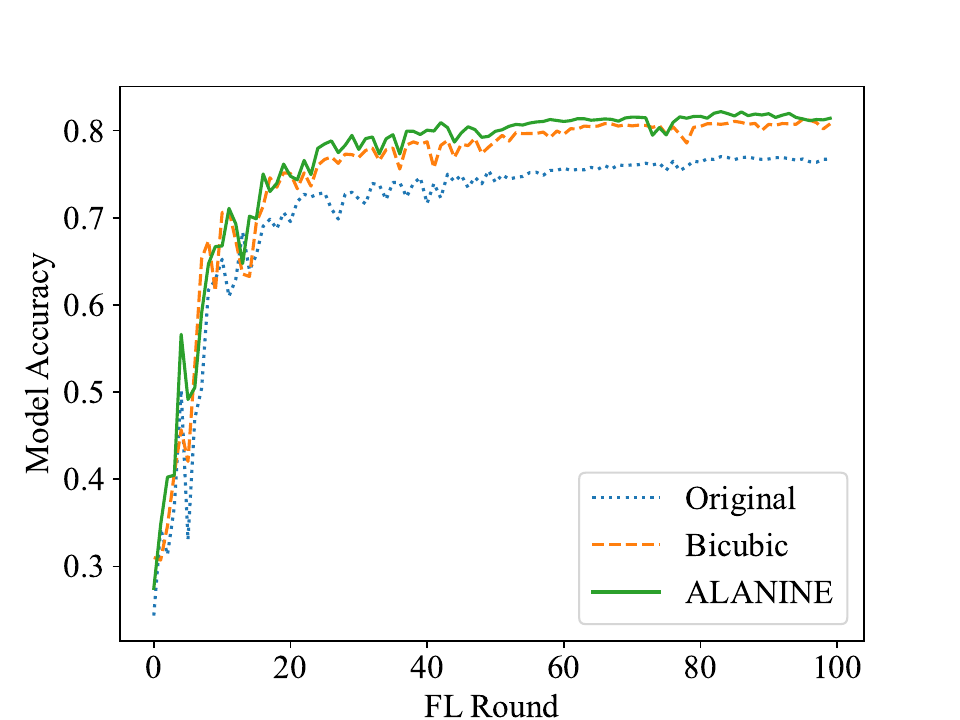}}
 \hspace{0in} 
\caption{ (a) and (b) respectively illustrate the model accuracy for FL rounds under $\times 2$ upscaling dataset with $\alpha = 0.3$, for local iterations of $1$ and $5$. (c), (d) depict the model accuracy for FL rounds under $\times 2$ upscaling dataset with $\alpha = 3$.}
 \label{fig5.1} 
\end{center}
\end{figure*}

\begin{figure*} [htb]
 \begin{center}
 \subfigure[]{
  \label{fig2a} 
  \includegraphics[width=4.2cm]{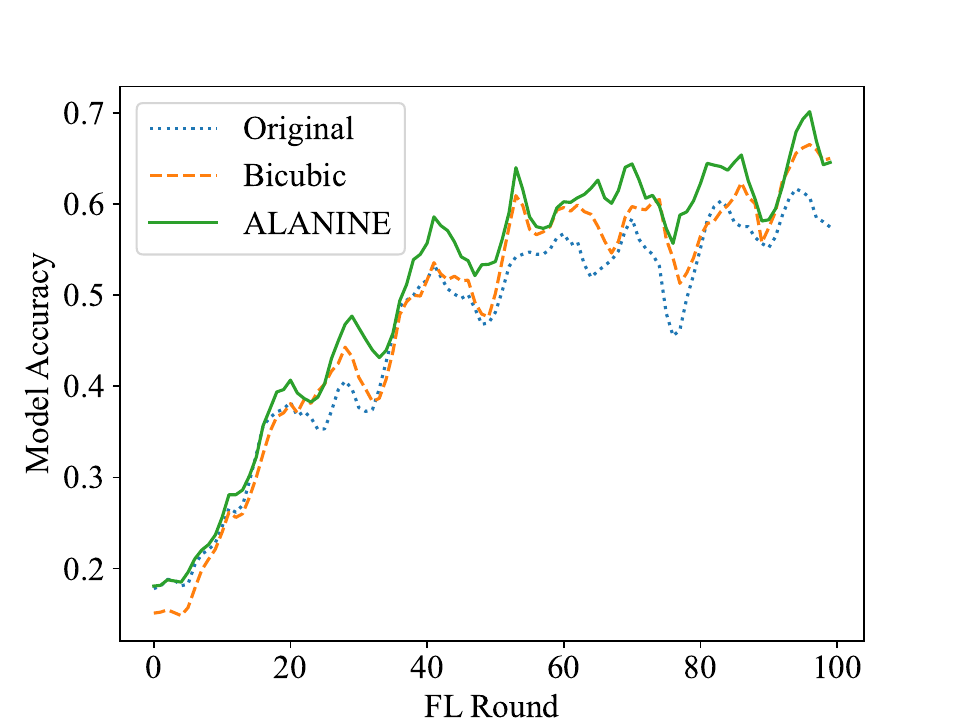}}
 \hspace{0in} 
 \subfigure[]{
  \label{fig2b} 
  \includegraphics[width=4.2cm]{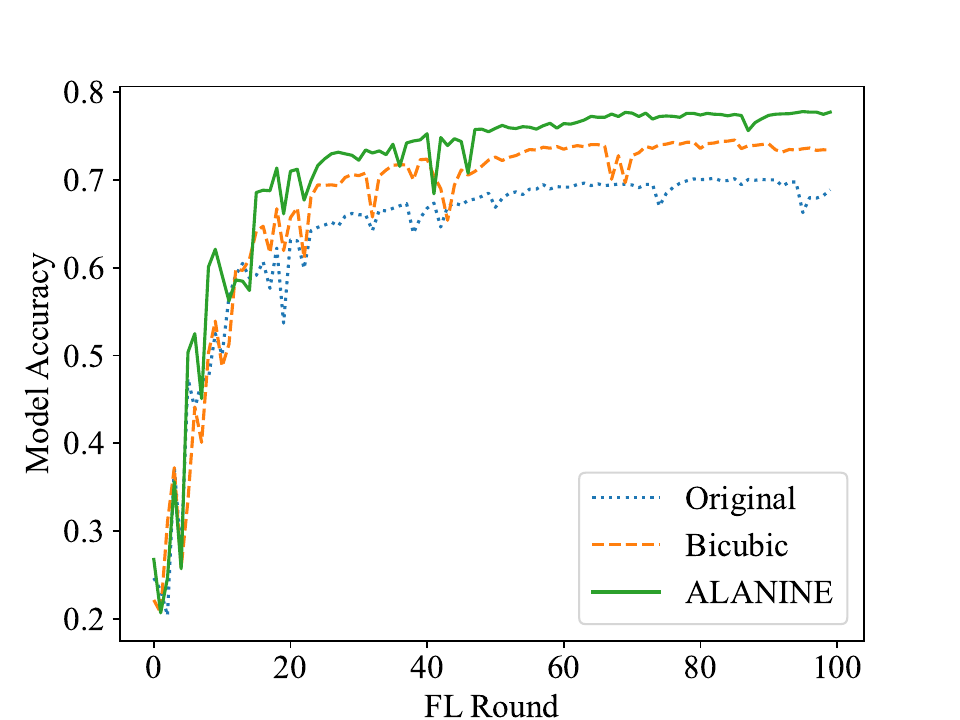}}
 \hspace{0in} 
 \subfigure[]{
  \label{fig2c} 
  \includegraphics[width=4.2cm]{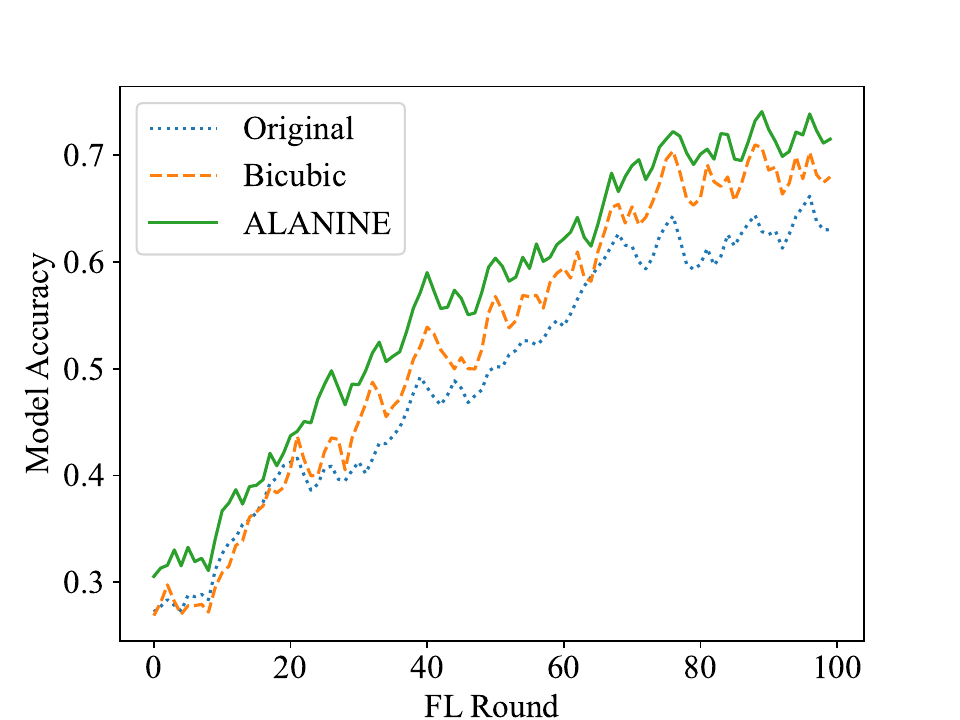}}
 \hspace{0in} 
  \subfigure[]{
  \label{fig2d} 
  \includegraphics[width=4.2cm]{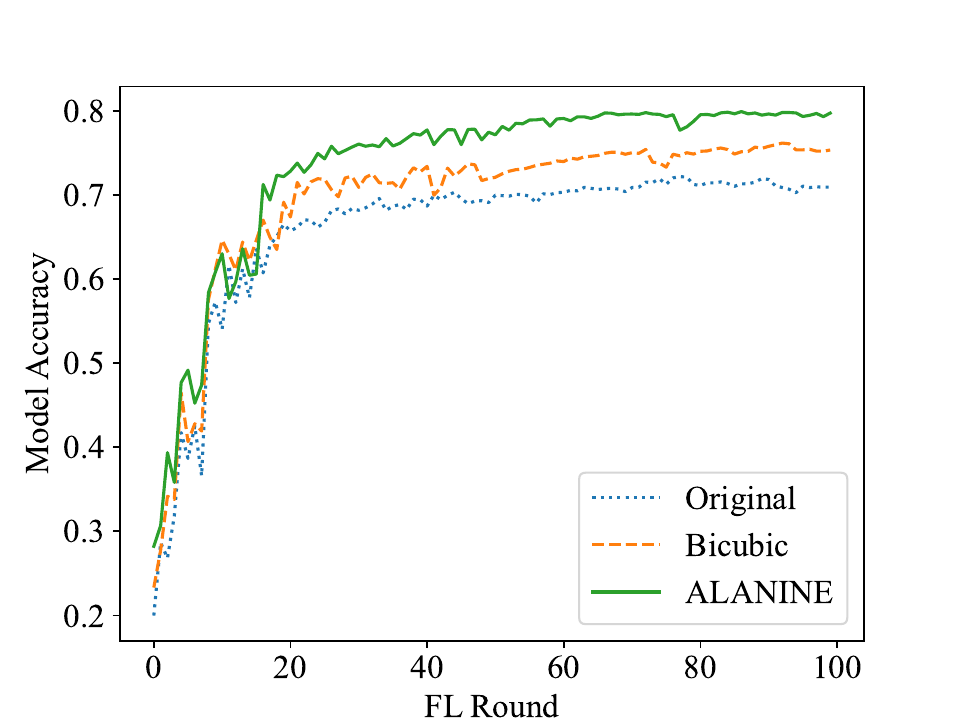}}
 \hspace{0in} 
\caption{ (a) and (b) demonstrate the model accuracy across FL rounds for $\times 4$ upscaling dataset when $\alpha = 0.3$, with local iteration $1$ and $5$, respectively. (c), (d) represent the model accuracy for different FL rounds under $\times 4$ upscaling dataset with $\alpha = 3$.}
 \label{fig5.2} 
\end{center}
\end{figure*}

\begin{itemize}

\item  FederatedAveraging (\emph{FedAvg}) \cite{FedAvg}: This algorithm implements a distributed learning approach where multiple clients collaboratively train a shared model while keeping their data locally.
\item  Decentralized sparse training based PFL (\emph{DisPFL}) \cite{DisPFL}: This strategy features modified gossip averaging for sparse models, local training with fixed sparse masks, and a mask evolution process to customize models for each client data distribution.
\item  Communication-efficient DFL (\emph{GossipFL}) \cite{GossipFL}: This approach uses model sparsification and adaptive peer selection to enable efficient DFL, where clients exchange sparse models with single peers chosen dynamically based on available bandwidth.
\item  Decentralized FedAvg with momentum (\emph{DFedAvgM}) \cite{DFedAvgM}: In this method, FedAvg is extended to a decentralized setting, where clients connected in a graph perform local SGD with momentum updates, then average models with neighbors using a gossip-based scheme.

\end{itemize}

\subsubsection{The performance of ALANINE with different data distributions and local iterations}

Fig. \ref{fig5.1} illustrates the performance of different schemes in terms of model accuracy across FL rounds for $\times 2$ upscaling task. The x-axis and y-axis represent the FL round and model accuracy, respectively. We evaluate three distinct image processing schemes on $\times 2$ upscaling dataset, which include the original image, Bicubic interpolation, and ALANINE SR. These methods are compared across varying FL rounds and local iterations to assess their performance and efficacy in the context of distributed learning scenarios. Fig. \ref{fig1a} and Fig. \ref{fig1b} correspond to $\alpha = 0.3$ with local iterations of 1 and 5, respectively, while Fig. \ref{fig1c} and Fig. \ref{fig1d} show results for $\alpha = 3$ under the same iteration settings.

In Fig. \ref{fig1a} and Fig. \ref{fig1b}, three schemes demonstrate a positive trend in accuracy as the number of FL rounds and local iterations increase. The primary distinction lies in the varying number of local iterations, which results in significant differences in the accuracy curve behavior across FL rounds. In Fig. \ref{fig1a}, accuracy fluctuations are more pronounced, whereas Fig. \ref{fig1b} demonstrates enhanced stability in accuracy during the later stages of training. Our proposed ALANINE SR method for image preprocessing yields superior classification model accuracy compared to the other two approaches, particularly in the later rounds of FL. While performance is comparable across all methods in the early stages, as the learning process progresses, training with the ALANINE SR method exhibits a marked improvement. Consistent with the aforementioned training results, Fig. \ref{fig1c} and Fig. \ref{fig1d} exhibit an ascending trend in image classification performance across all three image processing methods. The higher value of $\alpha$ results in a more uniform data distribution, which leads to improved overall model training results. However, this may also lead to increased fluctuations in accuracy during the training process. This observation indicates that the distribution of data significantly influences the training accuracy of FL models.

Building upon the analysis of the $\times 2$ upscaling scenario, we extend our investigation to examine the performance of our proposed methods in a more challenging $\times 4$ upscaling context. Fig. \ref{fig5.2} presents a comparative analysis of model accuracy across FL rounds for this task. In this context, we maintain our comparative analysis of three image processing approaches. This consistent methodology enables a direct comparison with the $\times 2$ upscaling results while offering insights into the performance of each approach under more challenging conditions. The simulation parameters are consistent with those of the preceding case, with Fig. \ref{fig5.2} illustrating results for various combinations of $\alpha$ values and local iterations. This configuration allows us to assess the scalability and robustness of our approach under more demanding upscaling conditions.

\begin{figure} [t]
 \subfigure[]{
  \label{fig3a} 
  \includegraphics[width=4.3cm]{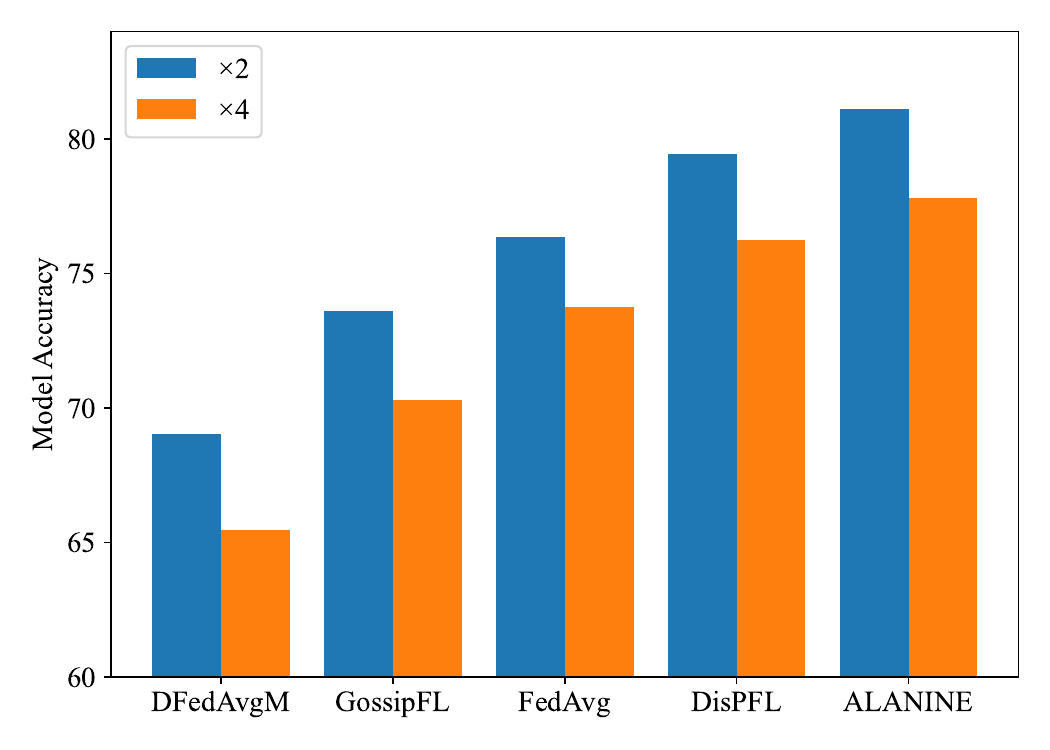}}
 \hspace{0in} 
 \subfigure[]{
  \label{fig3b} 
  \includegraphics[width=4.3cm]{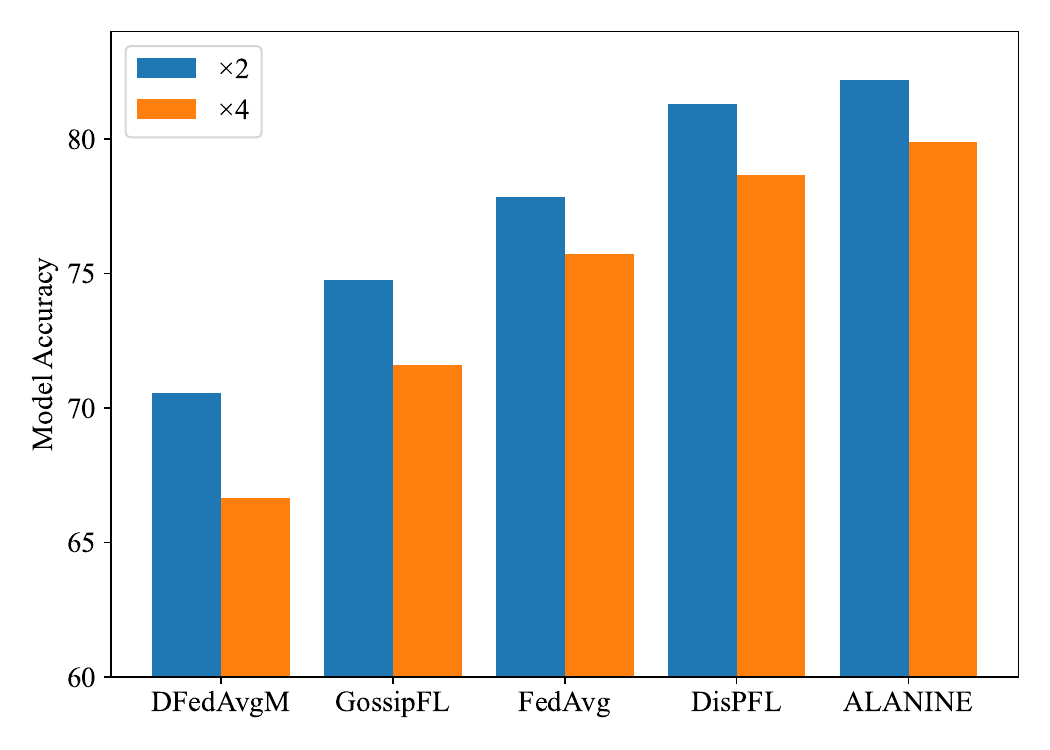}}
 \hspace{0in} 
 \caption{Comparative analysis of model accuracy across five schemes under varying Dirichlet parameters (a) $\alpha = 0.3$ and (b) $\alpha = 3$.} 
 \label{fig5.3} 
\end{figure}

Fig. \ref{fig5.2} reveal trends similar to those observed in the $\times 2$ upscaling scenario. All three preprocessing methods show an increase in accuracy as the number of FL rounds and local iterations grow. Notably, the ALANINE SR-processed images consistently maintain the highest accuracy throughout the training process, with a more pronounced advantage. However, it is important to note that the overall accuracy levels achieved for $\times 4$ upscaling are lower than those observed in the $\times 2$ case, reflecting the increased challenge of the task. This challenge can be attributed to the more significant disparities in image quality produced by different preprocessing methods in the $\times 4$ SR context. The upscaled images resulting from different preprocessing methods may exhibit varying degrees of clarity and detail retention, which leads to significant disparities in effective resolution. These differences in image quality have a pronounced impact on the model learning process and ultimate performance. Such findings highlight the increasing importance of sophisticated preprocessing techniques in high-resolution image classification tasks within FL environments, especially as the upscaling factor becomes more demanding. The superior performance of ALANINE SR in this context underscores its efficacy in preserving crucial image features, thus facilitating more effective model training and improved classification accuracy.

\subsubsection{The performance of ALANINE with other FL algorithms under different data distributions}

As depicted in Fig. \ref{fig5.3}, a comparative analysis of model accuracy across five FL schemes (DFedAvgM, GossipFL, FedAvg, DisPFL, and ALANINE) is presented under two Dirichlet parameters and two upscaling factors. ALANINE consistently outperforms other algorithms across different levels of data heterogeneity, with the performance gap more pronounced in scenarios where satellites have varying data acquisition capabilities, which results in non-uniform data distributions across the satellite constellation. Furthermore, as the upscaling factor decreases, the accuracy of each algorithm gradually improves. This observation indicates that the degree of data upscaling has a significant impact on model accuracy. Lower upscaling values correspond to the preservation of higher resolution image information, which contributes to enhanced model performance. Notably, ALANINE consistently outperforms other algorithms across different upscaling factors, which demonstrates its capacity to effectively utilize high-fidelity data. These results underscore ALANINE robustness and efficiency in handling diverse data distributions within FL scenarios.

\begin{figure} [t]
 \subfigure[]{
  \label{fig4a} 
  \includegraphics[width=4.3cm]{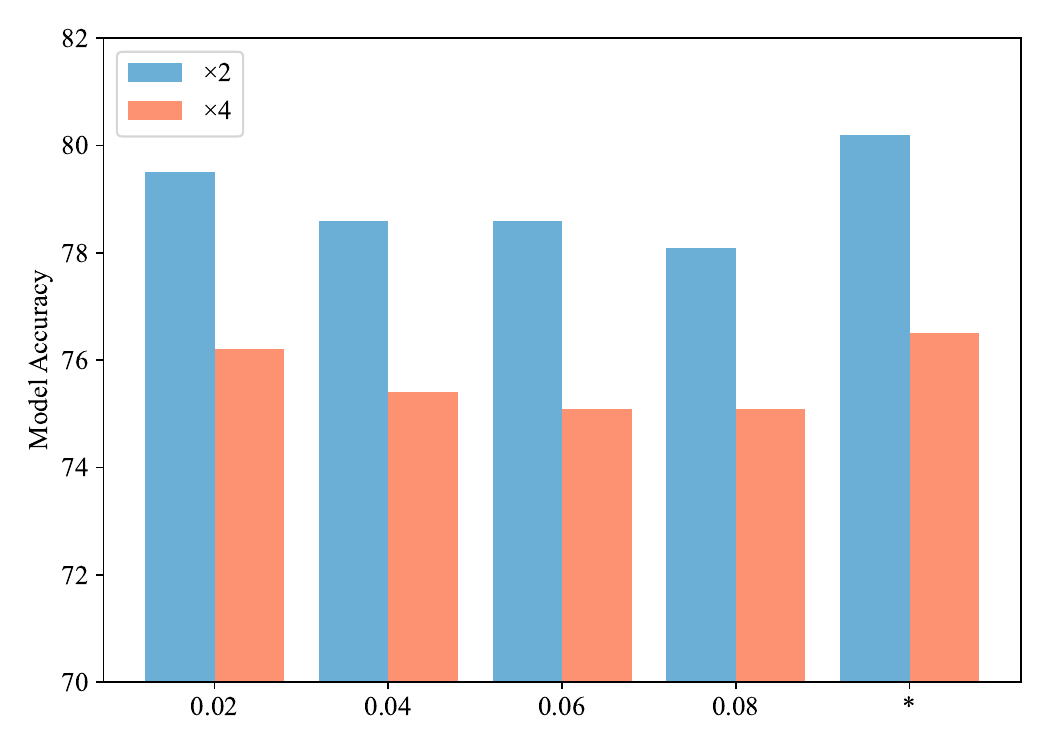}}
 \hspace{0in} 
 \subfigure[]{
  \label{fig4b} 
  \includegraphics[width=4.3cm]{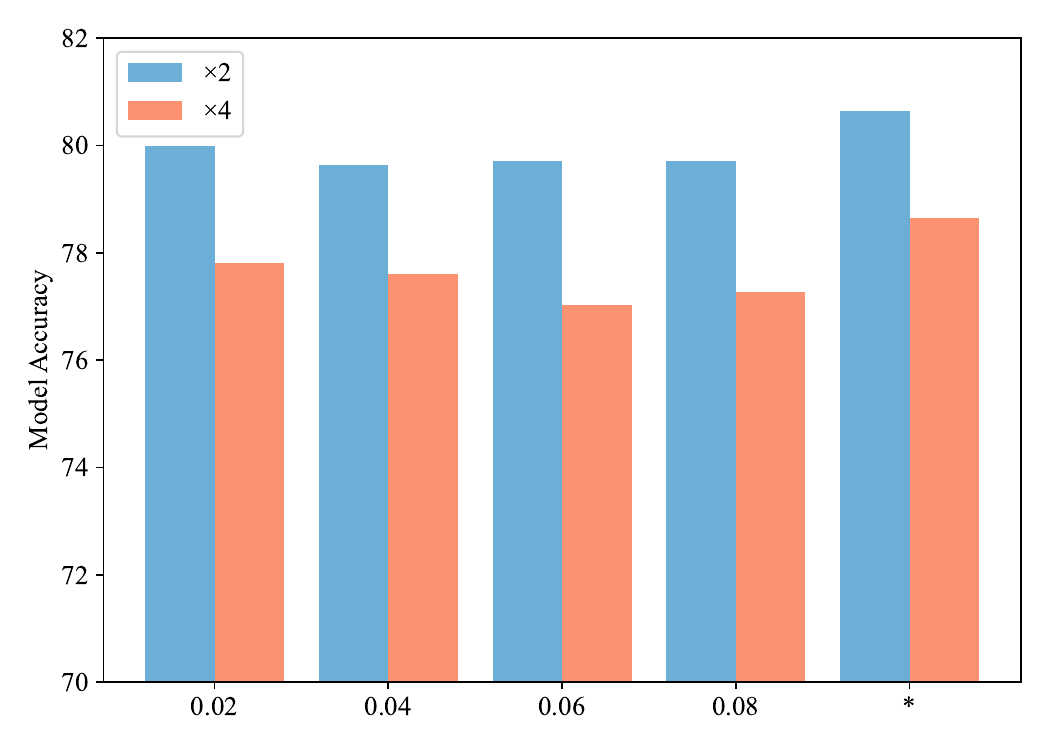}}
 \hspace{0in} 
 \caption{Comparative analysis of model accuracy across different $\varepsilon_c$ under varying Dirichlet parameters (a) $\alpha = 0.3$ and (b) $\alpha = 3$.} 
 \label{fig5.4} 
\end{figure}

\subsubsection{The performance of ALANINE under different pruning thresholds}

Fig.\ref{fig5.4} presents a comparative analysis of model accuracy across different $\varepsilon_c$ thresholds for two upscaling factors under varying Dirichlet parameters and elucidates the impact of pruning on model performance. The results demonstrate ALANINE efficacy across diverse data heterogeneity conditions, with $\alpha = 0.3$ and $\alpha = 3$ representing different degrees of data non-uniformity across the network. The $\ast$ represents our proposed ALANINE method, which maintains the initially set pruning sparsity level of $0.6$ without requiring further pruning. As $\varepsilon_c$ values increase, we observe a gradual decline in accuracy for both upscaling image types, which demonstrates the inverse relationship between $\varepsilon_c$ and model performance. Based on these results, we selected an $\varepsilon_c$ range around $0.02$, as it yields performance closest to that of $\ast$. This choice balances the trade-off between pruning effectiveness and maintaining model accuracy. Notably, the performance disparity between upscaling factors remains consistent across different data heterogeneity levels, which underscores ALANINE robustness and efficiency in FL contexts. 

\begin{figure}[t]
  \centering
  \includegraphics[width=0.4\textwidth]{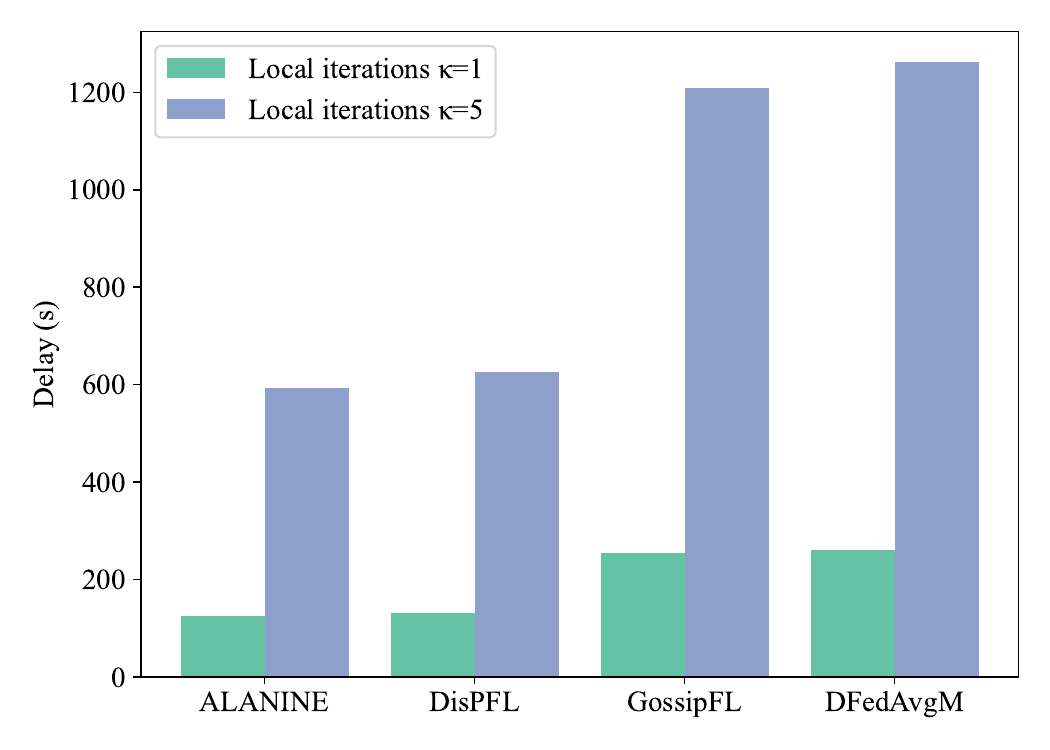}
  \caption{Comparative analysis of delay across different DFL.}
  \label{fig5.5}
\end{figure}

\subsubsection{The Delay of ALANINE with other DFL algorithms across different FL rounds}

Fig. \ref{fig5.5} illustrates the delay performance of four distinct DFL approaches under varying local iteration $\kappa$. As the number of local iterations increases from $1$ to $5$, all four DFL methods exhibit a significant increase in delay, with the magnitude of delay approximately quintupling across the board. The delay calculation is performed using the equation derived from \cite{FL4}. In our proposed method, this equation incorporates both computation and communication delays inherent in distributed learning systems. Notably, our proposed ALANINE algorithm demonstrates substantially lower delay compared to the other three methods under both local iteration scenarios. This superior performance can be primarily attributed to our innovative approach in model training, which incorporates model pruning and dynamic aggregation techniques. These strategies effectively reduce the model training size and enhance training efficiency. Using these advanced methods, ALANINE successfully mitigates the computational and communication overhead typically associated with distributed learning, resulting in a more efficient DFL process. 

\subsection{Summary of Results}

The evaluation of ALANINE demonstrates its superior performance across multiple aspects. In satellite image SR, ALANINE consistently outperforms existing techniques in both PSNR and SSIM metrics for different upscaling factors (Table \ref{tab3}). These improvements translate to tangible enhancements in image quality, particularly important for high-resolution satellite imagery analysis. Qualitative analysis also shows ALANINE achieves improved reconstruction of fine textures, smoother color transitions, and better preservation of structural details, which is essential for accurate interpretation of remote sensing data (Fig. \ref{fig:A}).

In the PFL evaluation, ALANINE exhibits higher model accuracy compared to other preprocessing methods across different data distributions and upscaling factors (Fig. \ref{fig5.1} and Fig. \ref{fig5.2}). This consistency in performance in varing conditions underscores the adaptability of ALANINE to the heterogeneity of real-world satellite data. ALANINE also consistently outperforms other FL algorithms under various conditions (Fig. \ref{fig5.3}). The framework ability to maintain high accuracy even with non-uniform data distributions highlights its potential for practical application in diverse satellite constellations. Furthermore, ALANINE can optimize balance pruning efficiency and model accuracy based on the settings of different pruning thresholds (Fig. \ref{fig5.4}).

Finally, ALANINE demonstrates superior efficiency in terms of delay (Fig. \ref{fig5.5}). When compared to other DFL approaches, ALANINE exhibits substantially lower delay under different local iteration scenarios. This efficiency gain is particularly significant in satellite networks where communication bandwidth is limited and computational resources are constrained. The improved performance is attributed to ALANINE innovative approach in model training, which incorporates model pruning and dynamic aggregation techniques, effectively reducing model training size and enhancing training efficiency. These optimizations not only improve processing speed but also contribute to more efficient utilization of satellite resources.

\section{CONCLUSION} {\label{conclusion}}
We present ALANINE, an innovative decentralized PFL framework designed for heterogeneous LEO satellite constellations. ALANINE addresses critical challenges in satellite-based data processing, namely data heterogeneity, varying image resolutions, and efficient on-orbit ML model training. The framework integrates DFL for satellite image SR, which enhances input data quality, and implements a personalized approach to account for unique satellite data characteristics. In addition, it employs advanced model pruning to optimize model complexity and transmission efficiency. The simulation results demonstrate ALANINE superior performance in on-orbit training of SR and PFL image processing models compared to traditional centralized approaches. This novel framework shows significant improvements in data acquisition efficiency, processes accuracy, and model adaptability to local satellite conditions. This research contributes to the advancement of Earth observation applications by enabling more efficient and accurate on-orbit data processing while addressing crucial issues of data privacy and communication bandwidth constraints in satellite networks.

\bibliography{sample}
\bibliographystyle{IEEEtran}

\end{document}